\documentclass[apjl]{emulateapj}
\usepackage{psfig,amsfonts,amsmath,graphicx,natbib,apjfonts,lscape}

\def\nod{\nodata}

\def\cfa{1}

\begin{document}

\title{A Short GRB ``No-Host'' Problem?  \\ Investigating Large
Progenitor Offsets for Short GRBs with Optical Afterglows}

\author{ 
E.~Berger\altaffilmark{\cfa}
}

\altaffiltext{\cfa}{Harvard-Smithsonian Center for Astrophysics, 60
Garden Street, Cambridge, MA 02138}

\begin{abstract} We investigate the afterglow properties and
large-scale environments of several short-duration gamma-ray bursts
(GRBs) with sub-arcsecond optical afterglow positions but no bright
coincident host galaxies.  The purpose of this joint study is to
robustly assess the possibility of significant offsets, a hallmark of
the compact object binary merger model.  Five such events exist in the
current sample of 20 short bursts with optical afterglows, and we find
that their optical, X-ray, and $\gamma$-ray emission are
systematically fainter.  These differences may be due to lower
circumburst densities (by about an order of magnitude), to higher
redshifts (by $\Delta z\approx 0.5-1$), or to lower energies (by about
a factor of 3), although in the standard GRB model the smaller
$\gamma$-ray fluences cannot be explained by lower densities.  To
study the large-scale environments we use deep optical observations to
place limits on underlying hosts and to determine probabilities of
chance coincidence for galaxies near each burst.  In 4 of the 5 cases
the lowest probabilities of chance coincidence ($P(<\delta R)\sim
0.1$) are associated with bright galaxies at separations of $\delta
R\sim 10\arcsec$, while somewhat higher probabilities of chance
coincidence are associated with faint galaxies at separations of $\sim
2\arcsec$.  By measuring redshifts for the brighter galaxies in three
cases ($z=0.111,\,0.473,\,0.403$) we find physical offsets of $\approx
30-75$ kpc, while for the faint hosts the assumption of $z\gtrsim 1$
leads to offsets of $\sim 15$ kpc.  Alternatively, the limits at the
burst positions ($\gtrsim 26$ mag) can be explained by typical short
GRB host galaxies ($L\approx 0.1-1$ L$^*$) at $z\gtrsim 2$.  Thus, two
possibilities exist: (i) $\sim 1/4$ of short GRBs explode $\sim 50$
kpc or $\sim 15$ kpc from the centers of $z\sim 0.3$ or $z\gtrsim 1$
galaxies, respectively, and have fainter afterglows due to the
resulting lower densities; or (ii) $\sim 1/4$ of short GRBs occur at
$z\gtrsim 2$ and have fainter afterglows due to their higher
redshifts.  The high redshift scenario leads to a bimodal redshift
distribution, with peaks at $z\sim 0.5$ and $z\sim 2$.  The large
offset scenario leads to an offset distribution that is well-matched
by theoretical predictions of NS-NS/NS-BH binary kicks, or by a hybrid
population with globular cluster NS-NS binaries at large offsets and
primordial binaries at offsets of $\lesssim 10$ kpc (indicative of
negligible kicks).  Deeper constraints on any coincident galaxies to
$\gtrsim 28$ mag (using the {\it Hubble Space Telescope}) will allow
us to better exclude the high-redshift scenario.  \end{abstract}


\section{Introduction}
\label{sec:into}

The bimodality of gamma-ray burst (GRB) durations \citep{kmf+93} is
indicative of separate progenitor populations for long- and
short-duration GRBs.  While direct observational support exists for
the massive star origin of long GRBs (e.g., \citealt{wb06}), the most
popular progenitor model for short GRBs is the coalescence of compact
object binaries with neutron star and/or black hole constituents
(NS-NS/NS-BH; \citealt{elp+89,pac91,npp92}).  One of the key
predictions of this model is that some systems will experience large
velocity kicks at birth, leading to eventual mergers well outside of
the host galaxies, in galactic halos and the intergalactic medium
\citep{bsp99,fwh99,bpb+06}.  These models predict that $10-20\%$ of
all mergers will occur at offsets of $\gtrsim 20$ kpc for Milky Way
mass galaxies.  In such environments the resulting afterglow emission
is expected to be fainter than for bursts occurring in coincidence
with their host galaxies due to the low ambient density
\citep{pkn01,pb02}.

A subset of NS-NS binaries ($\sim 10-30\%$) may be formed dynamically
in globular clusters \citep{gpm06}, leading to possible large offsets
and low ambient densities despite an absence of kicks.  The
distribution of offsets for such binary systems can in principle be
calculated from the spatial distribution of globular clusters, and
initial predictions are that $50-90\%$ of these systems will have
offsets of $\gtrsim 20$ kpc (depending on the mass of the galaxy;
\citealt{sdc+10}).  An additional expectation is that
dynamically-formed binaries will be heavily skewed to lower redshifts
due to the additional time delay between the formation and
core-collapse of the globular clusters \citep{hgw+06,scc+08}.

Other progenitor systems for short GRBs have also been proposed,
including young magnetars \citep{td95}, accretion-induced collapse
(AIC) of neutron stars \citep{qwc+98}, and delayed magnetar formation
through binary white dwarf mergers or white dwarf AIC
\citep{lwc+06,mqt08}.  These models are partially motivated by
observations that cannot be easily accommodated in the standard NS-NS
merger model, such as extended soft $\gamma$-ray emission on
timescales of $\sim 100$ s (e.g., \citealt{vlr+05,mqt08,pmg+09}), or
by phenomena such as short-duration giant flares from soft
$\gamma$-ray repeaters (e.g., \citealt{hbs+05,pbg+05,tcl+05,ngp+06}).
The general expectation is that these alternative progenitors will not
experience kicks, and will therefore lead to bursts in coincidence
with host galaxies.

The detection of short GRB afterglows starting in mid-2005 provided an
opportunity to investigate the various progenitor models through a
range of observational tests: the redshift distribution
\citep{bfp+07,gno+08}, the host galaxy demographics \citep{ber09}, the
afterglow properties \citep{ber07,gbb+08,kkz+08,nfp+09}, and perhaps
most importantly, their locations relative to the host galaxies
\citep{fbf10}.  As of mid-2010, X-ray and optical afterglows have been
detected from 40 and 20 short GRBs, respectively, with the latter
sample providing accurate sub-arcsecond positions.  Of these 20
events, 15 directly coincide with host galaxies with a wide
distribution of apparent magnitudes, and redshifts of $z\approx 0.2-1$
or beyond (e.g., \citealt{bfp+07,ber09,dmc+09}).  However, the
remaining five events\footnotemark\footnotetext{These are GRBs 061201:
\citet{gcn5952,sdp+07,fbf10}; 070809: \citet{gcn6739,gcn7889}; 080503:
\citet{pmg+09}; 090305: \citet{gcn8933,gcn8934}; and 090515:
\citet{row+10}.} do not appear to coincide with galaxies, and
therefore provide an opportunity to assess the possibility of large
progenitor offsets, and to test the validity of the NS-NS merger
models.

Significant offsets have been claimed previously, in particular for
GRBs 050509b and 060502b with projected offsets of $39\pm 13$ and
$73\pm 19$ kpc, respectively \citep{bpp+06,bpc+07}.  However, in both
cases only X-ray positions are available ($3.5\arcsec$ and
$4.4\arcsec$ radius, respectively), and the error circles contain
several galaxies consistent with a negligible offset
\citep{bfp+07,bpc+07}.  Moreover, in the case of GRB\,050509b the
X-ray error circle intersected the outer regions of the host, raising
the possibility that the progenitor system was formed in, rather than
kicked to, the outskirts of the host.  This possibility raises a
crucial point, namely that a substantial physical offset from the
center of the host does not necessarily point to a progenitor kick if
the burst still closely coincides with the host light distribution
\citep{fbf10}.  An illustrative example of this point is GRB\,071227
whose optical afterglow position coincides with the outskirts of
edge-on disk galaxies, with an offsets of about 15 kpc from the host
center \citep{dmc+09,fbf10}.

Large offsets have also been speculated in a few cases with precise
optical afterglow positions (GRBs 061201, 070809, and 080503;
\citealt{sdp+07,gcn7889,pmg+09}).  However, these claims have not been
investigated systematically, mainly because they were treated on a
case-by-case basis, with probabilistic arguments that prevented
conclusive associations.  These cases, combined with the ambiguity
inherent to X-ray positions, demonstrate that bursts with optical
afterglows are essential for reaching any robust conclusions about
progenitor offsets (due to kicks and/or a globular cluster origin).

Here we present the first systematic study of short GRBs with optical
afterglows and no coincident hosts, which combines their afterglow
properties with the large-scale environments.  The purpose of this
study is to statistically assess the possibility of offsets and to
compare this with alternative explanations (e.g., a high redshift
origin).  To achieve this goal we set our study in the broader context
of short GRBs that have optical afterglows {\it and} coincident hosts,
as well as short GRBs with only X-ray positions.  As we demonstrate
throughout the paper, such a combined study is essential since offsets
or high redshifts are expected to jointly affect both the afterglow
properties and the large-scale environments.  Our study also provides
a robust statistical assessment of {\it a posteriori} chance
coincidence probabilities, and the expected number of spurious
associations, which cannot be properly assessed for individual bursts.

The paper is organized as follows.  In \S\ref{sec:obs} we present deep
optical observations of the environments of GRBs 061201, 070809,
080503, 090305, and 090515, as well as spectroscopic observations of
bright galaxies near the burst positions for GRBs 061201, 070809, and
090515.  We study trends in the afterglow and prompt emission
properties of bursts with and without coincident hosts in
\S\ref{sec:ag}, and determine {\it a posteriori} probabilities of
chance coincidence as a function of projected angular offset for
galaxies near the position of each burst in \S\ref{sec:hosts}.  We
also determine projected physical and host-normalized offsets, and use
these in conjunction with the afterglow and prompt emission properties
to address two scenarios for the short bursts with optical emission
and no coincident bright hosts: (i) an origin in faint galaxies at
$z\gtrsim 2$ (\S\ref{sec:highz}), or (ii) substantial offsets from
galaxies at $z\sim 0.1-0.5$ or $z\gtrsim 1$ (\S\ref{sec:kick}).
Finally, in \S\ref{sec:disc} we present the offset distribution of all
short GRBs with optical afterglows (in the context of scenario ii),
and compare this distribution with predictions for NS-NS kicks and
dynamically-formed NS-NS binaries.  We further investigate whether the
circumburst densities that are required for the measured optical
magnitudes can be accommodated with a halo or IGM origin.  We draw
conclusions about the progenitors of short GRBs from our systematic
study in \S\ref{sec:conc}.

\section{Short GRB Sample and Observations}
\label{sec:obs}

We include in this investigation all 20 short GRBs with optical
afterglow detections as of June 2010.  This is the full subset of
events for which sub-arcsecond positions are available.  We stress
that this sample represents only about 1/3 of all short GRBs
discovered to date, and about 1/2 of the sample with X-ray afterglow
detections.  Thus, it is not a complete sample of short GRBs, but it
is the only subset for which we can investigate the possibility of
large offsets with meaningful statistical significance.  As we
demonstrate in \S\ref{sec:disc}, we do not expect this sample to be
strongly biased with respect to circumburst density, at least for
$n\gtrsim 10^{-5}$ cm$^{-3}$.  The properties of the 20 short GRBs, as
well as events with deep optical limits, are summarized in
Table~\ref{tab:shbs}.  As can be inferred from the Table, some of the
20 events with only X-ray detections do not have optical follow-up
observations, suggesting that the sample with optical afterglows may
be largely representative.

For the purpose of our investigation we define three sub-samples that
will be used throughout the paper: (i) {\it Sample 1}: short GRBs with
detected afterglows and coincident host galaxies (15 bursts); (ii)
{\it Sample 2}: the 5 short bursts with detected optical afterglows
and no bright coincident hosts; and (iii) {\it Sample 3}: short GRBs
with detected X-ray afterglows (from the {\it Swift} X-ray Telescope)
but no optical detections despite rapid follow-up observations (11
bursts).

\subsection{Optical Imaging}

For the bursts in {\it Sample 2} we use deep space- and ground-based
optical observations to place limits on the brightness of underlying
galaxies, and to assess the probability of chance coincidence for
nearby galaxies.  For GRB\,061201 we use {\it Hubble Space Telescope}
(HST) observations obtained with the Advanced Camera for Surveys (ACS)
in the F814W filter, with a total exposure time of 2244 s
\citep{fbf10}.  For GRB\,070809 we use $r$-band observations obtained
on 2008 January 14 UT with the Low Dispersion Survey Spectrograph
(LDSS3) mounted on the Magellan/Clay 6.5-m telescope, with a total
exposure time of 1500 s.  For GRB\,080503 we use HST Wide-Field
Planetary Camera 2 (WFPC2) observations in the F606W filter from 2009
January 30 UT, with a total exposure time of 4000 s \citep{pmg+09}.
We also use the limits on a coincident host inferred from a deeper
stack of HST observations by \citet{pmg+09}, with $m_{\rm AB}({\rm
F606W})\gtrsim 28.5$ mag.  For GRB\,090305 we use LDSS3 $r$-band
observations obtained on 2010 May 8 UT with a total exposure time of
2400 s.  Finally, for GRB\,090515 we use $r$-band observations
obtained with the Gemini Multi-Object Spectrograph (GMOS) mounted on
the Gemini-North 8-m telescope from 2009 May 15 UT with a total
exposure time of 1800 s.

\begin{figure}
%\epsscale{1}
%\plotone{fig1.eps}
\centering
\includegraphics[angle=0,width=5.0in]{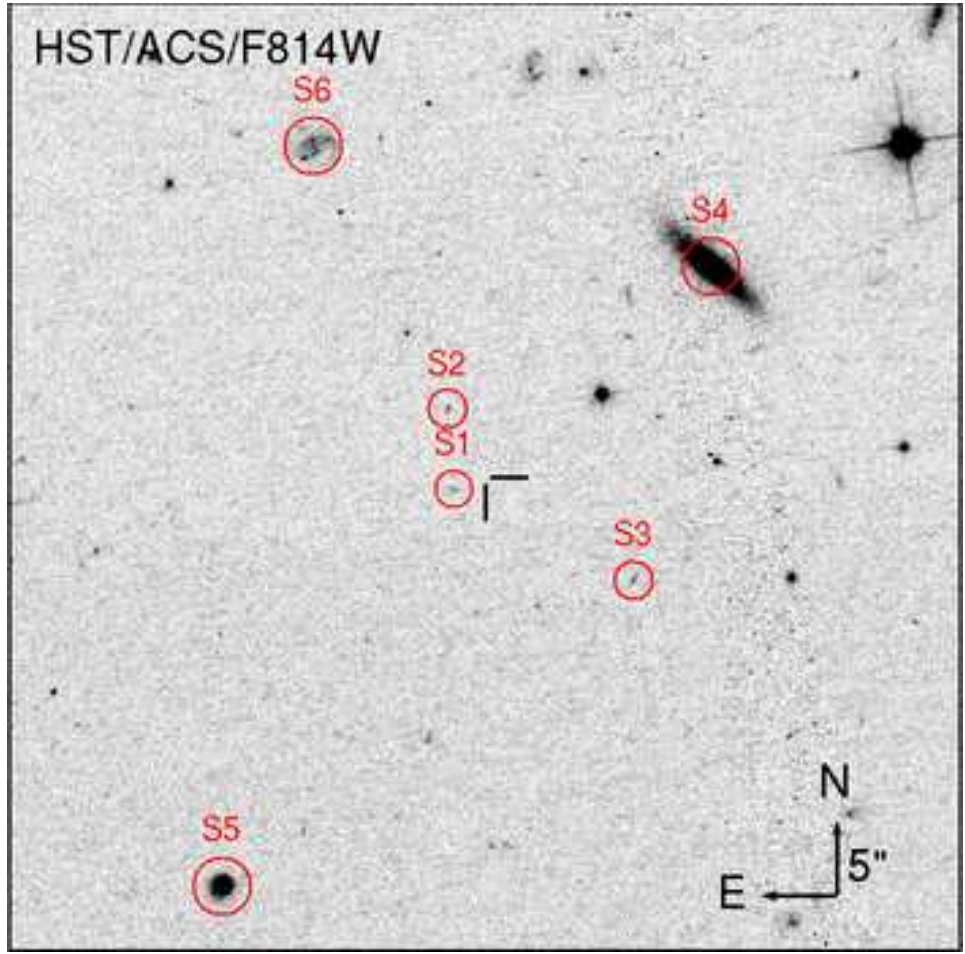}
\caption{{\it HST}/ACS/F814W image of the location of GRB\,061201.
Galaxies near the position of the optical afterglow (cross-hairs) are
marked.
\label{fig:061201}} 
\end{figure}

\begin{figure}
%\epsscale{1}
%\plottwo{070809_field_1_small.eps}{070809_field_2_small.eps}
\centering
\includegraphics[angle=0,width=3in]{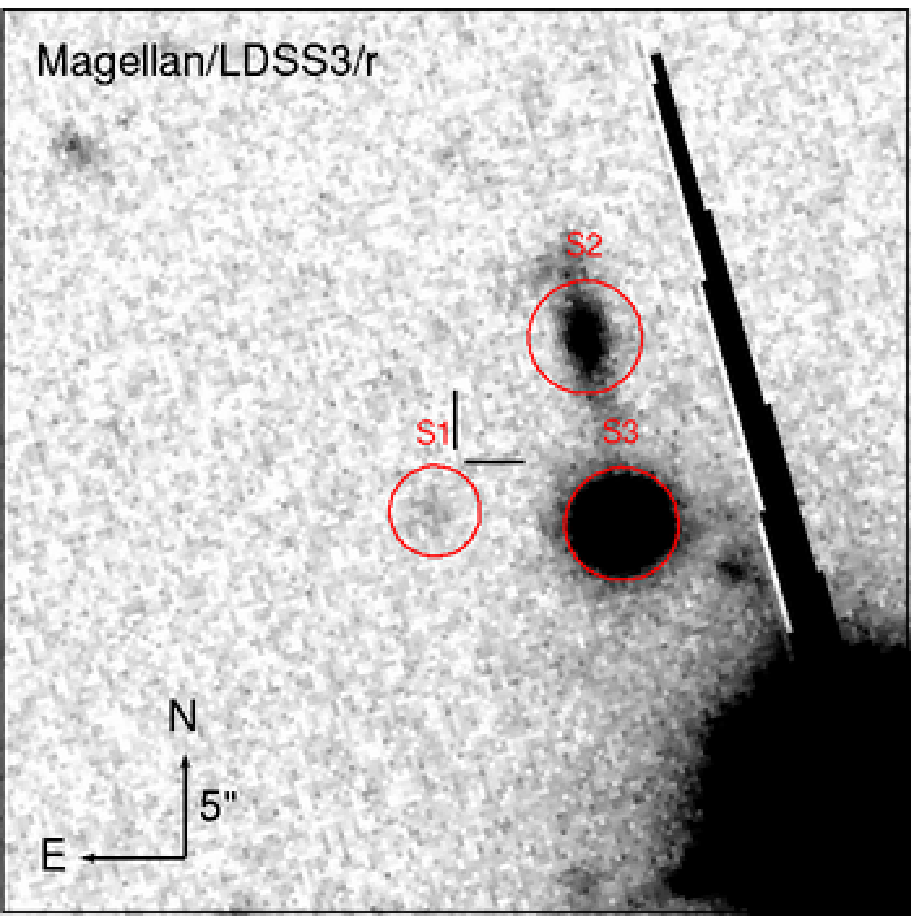}
\includegraphics[angle=0,width=3in]{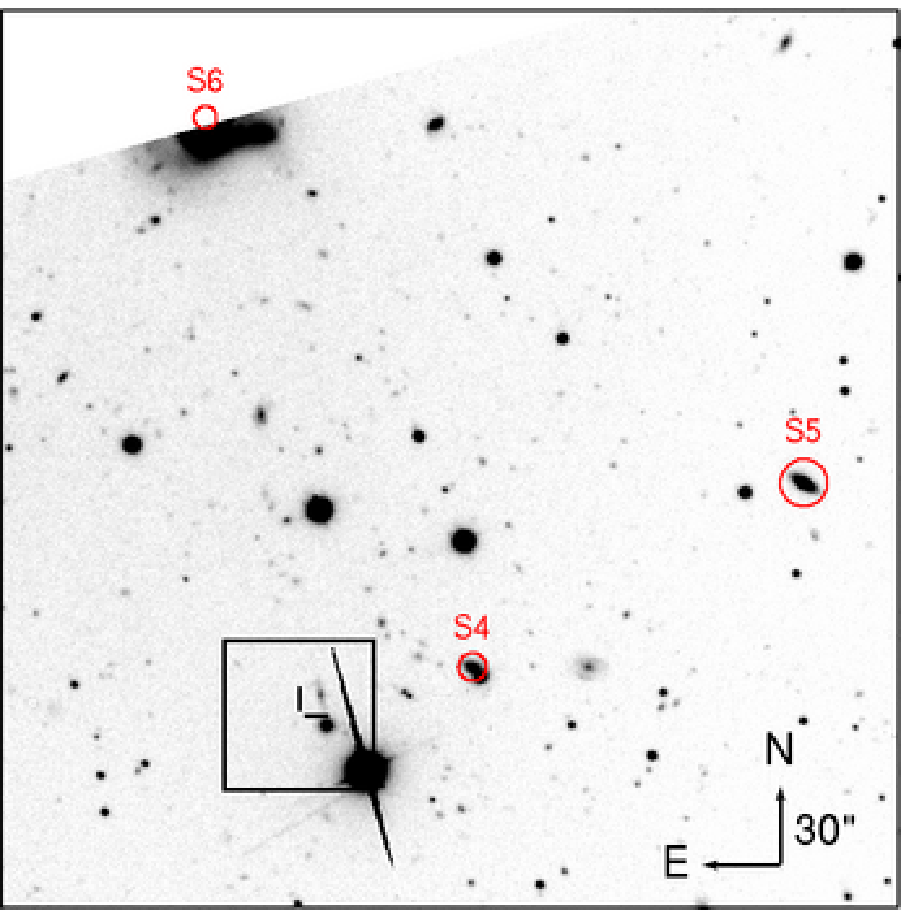}
\caption{Magellan/LDSS3 $r$-band images of the location of
GRB\,070809.  Galaxies near the position of the optical afterglow
(cross-hairs) are marked.  
\label{fig:070809}} 
\end{figure}

\begin{figure}
%\epsscale{1}
%\plotone{080503_field_small.eps}
\centering
\includegraphics[angle=0,width=5.0in]{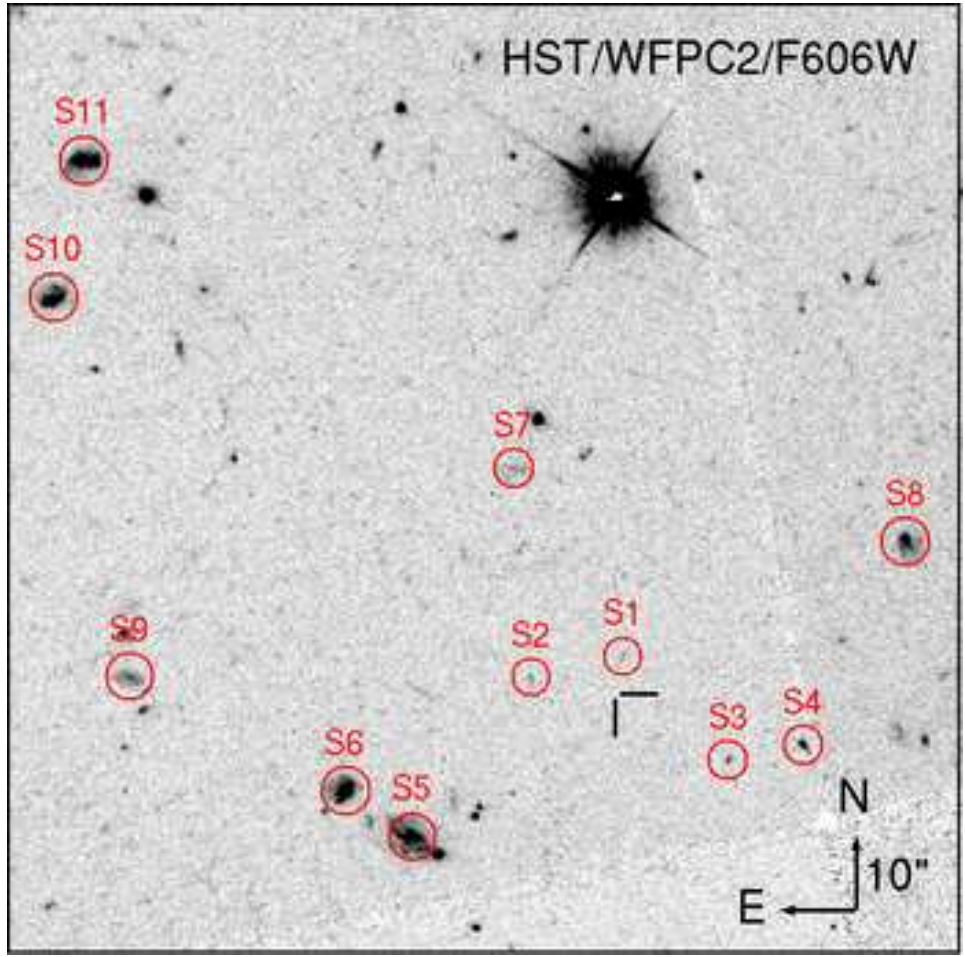}
\caption{{\it HST}/WFPC2/F606W image of the location of GRB\,080503.
Galaxies near the position of the optical afterglow (cross-hairs) are
marked.  A faint galaxy at a separation of only $0.8\arcsec$ was found
by \citet{pmg+09} based on a deeper stack of HST/WFPC2 observations.
These authors also find that the galaxy marked ``S5'' is located at
$z=0.561$, leading to a physical offset of 85 kpc.
\label{fig:080503}} 
\end{figure}

\begin{figure}
%\epsscale{1}
%\plottwo{090305_field_1_small.eps}{090305_field_2_small.eps}
\centering
\includegraphics[angle=0,width=3in]{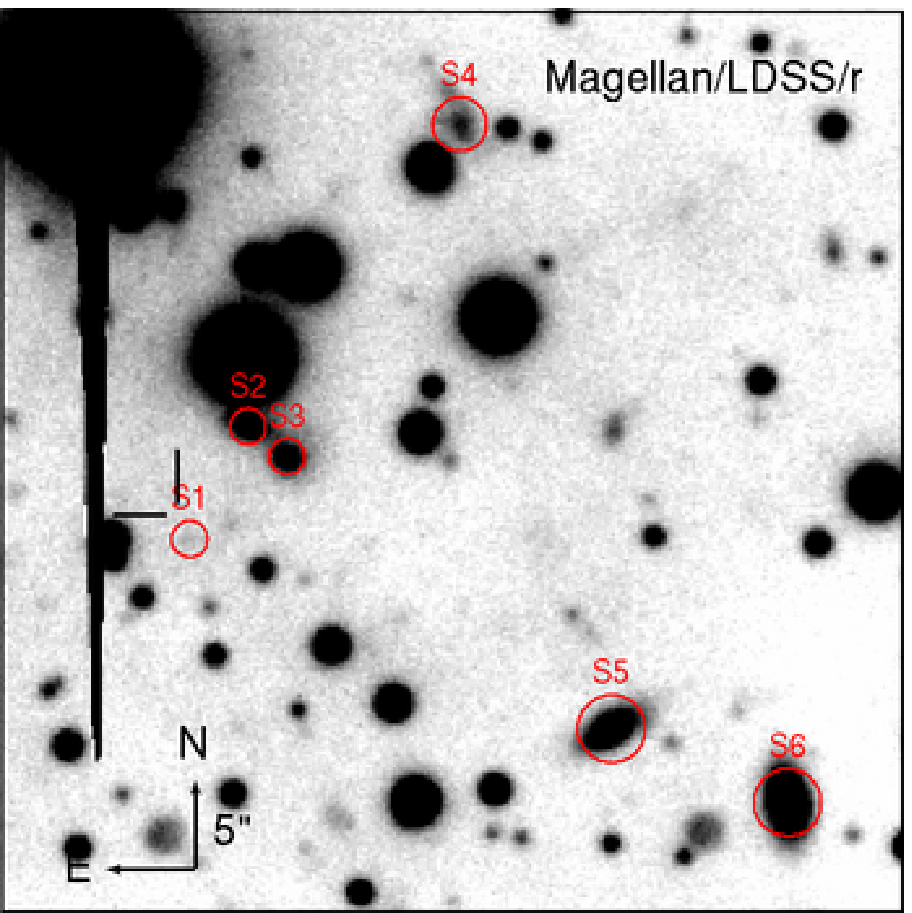}
\includegraphics[angle=0,width=3in]{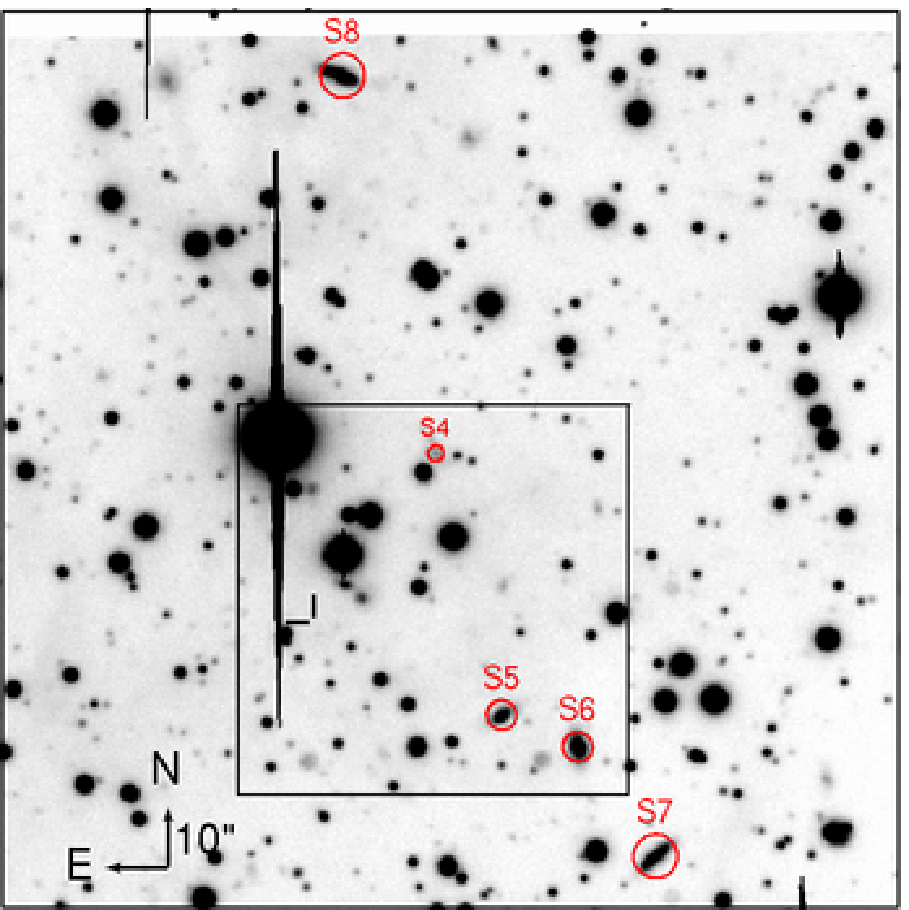}
\caption{Magellan/LDSS3 $r$-band images of the location of
GRB\,090305.  Galaxies near the position of the optical afterglow
(cross-hairs) are marked.  
\label{fig:090305}} 
\end{figure}

\begin{figure}
%\epsscale{1}
%\plottwo{090515_field_1_small.eps}{090515_field_2_small.eps}
\centering
\includegraphics[angle=0,width=3in]{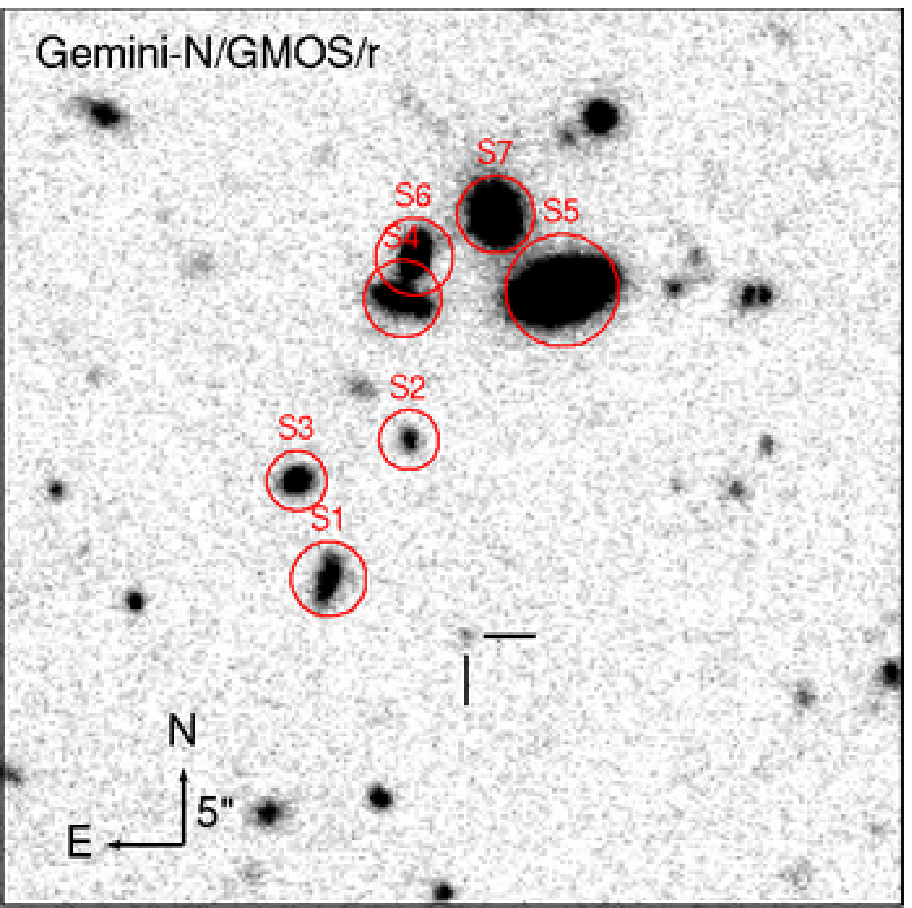}
\includegraphics[angle=0,width=3in]{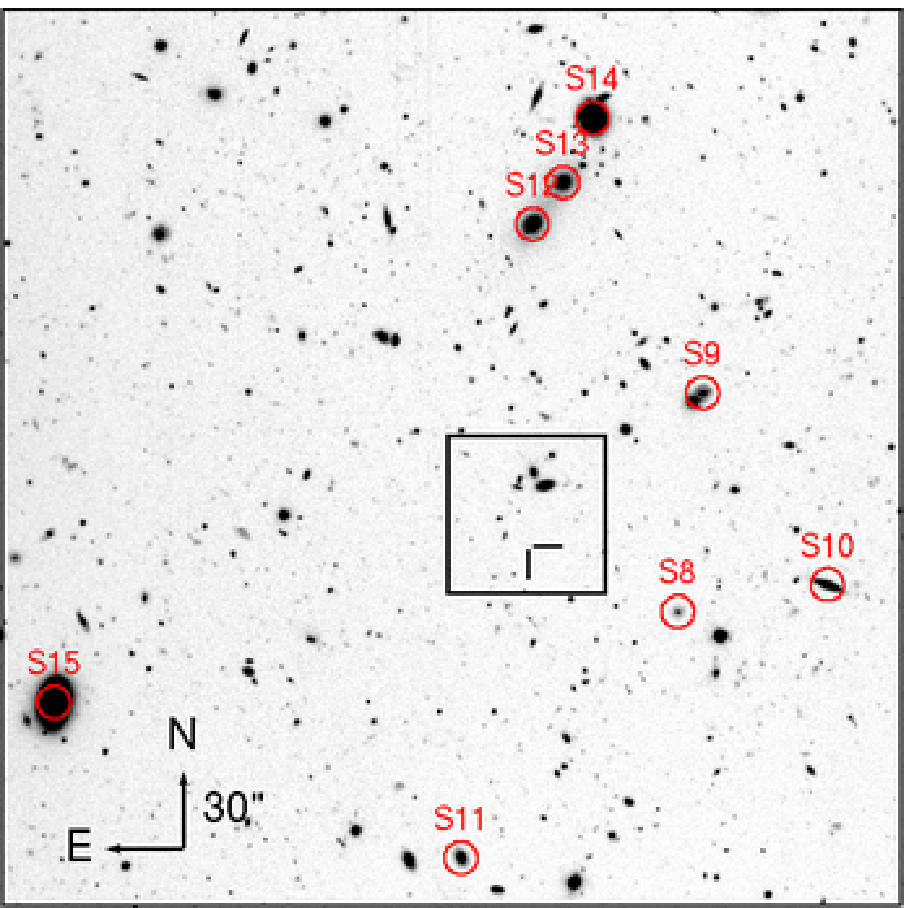}
\caption{Gemini-North/GMOS $r$-band images of the location of
GRB\,090515.  Galaxies near the position of the optical afterglow
(cross-hairs) are marked.  Note that the object coincident with the
cross-hairs is the optical afterglow.
\label{fig:090515}} 
\end{figure}

The ground-based observations were reduced and analyzed using standard
routines in IRAF.  The analysis of the HST observations is detailed in
\citep{fbf10}.  The limiting magnitudes for all five observations are
listed in Table~\ref{tab:nohost}, and images of the five fields are
shown in Figures~\ref{fig:061201}--\ref{fig:090515}.

\subsection{Optical Spectroscopy}

\begin{figure}
\epsscale{1}
\plotone{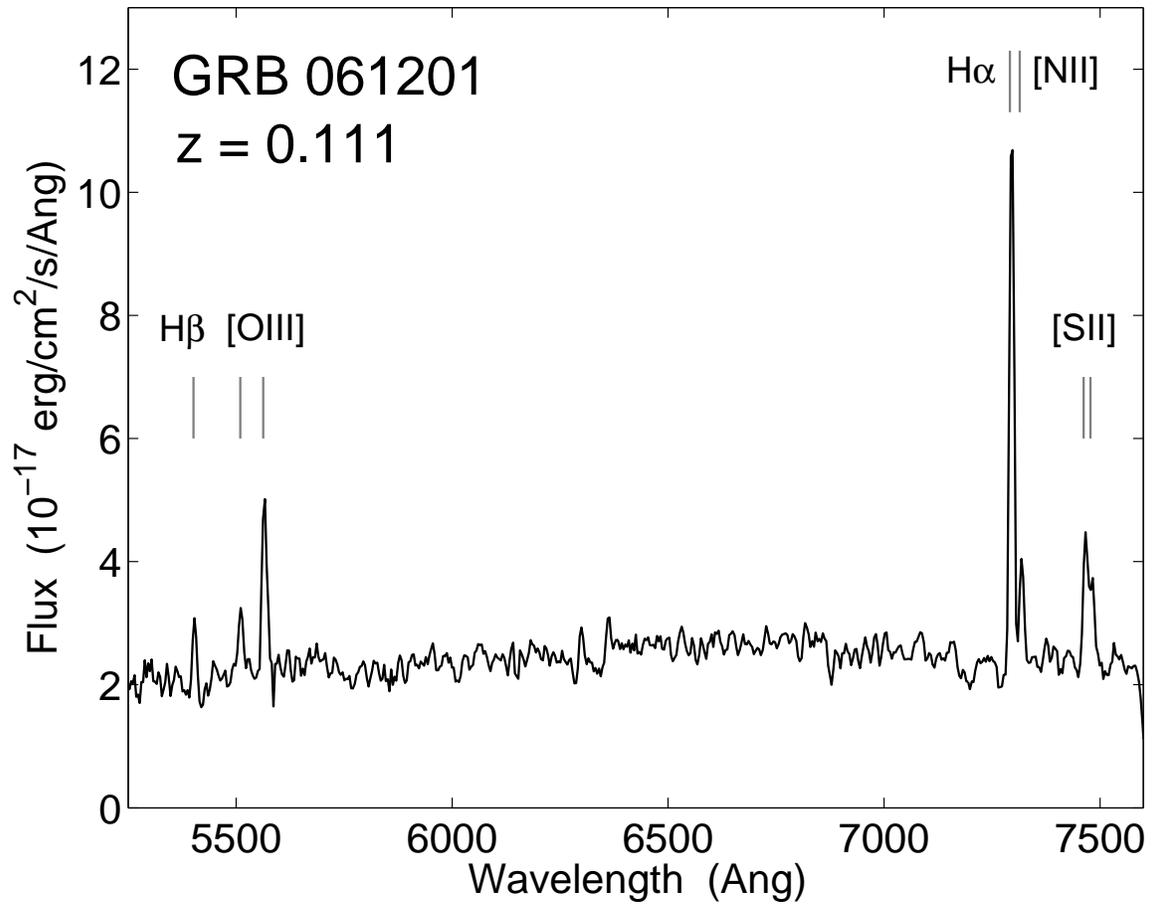}
\caption{Magellan/LDSS3 spectrum of the galaxy with the lowest
probability of chance coincidence near the position of GRB\,061201.
This galaxy is marked ``S4'' in Figure~\ref{fig:061201}.  It has a
redshift of $z=0.111$ and it is undergoing active star formation
\citep{gcn5952,sdp+07,fbf10}.
\label{fig:061201spec}} 
\end{figure}

In addition to the imaging observations, we obtained spectroscopic
observations of galaxies near the positions of GRBs 061201, 070809,
and 090515.  Our observations of a galaxy located $16.3\arcsec$ from
the afterglow position of GRB\,061201 (marked ``S4'' in
Figure~\ref{fig:061201}) revealed a star forming galaxy at $z=0.111$
\citep{sdp+07,fbf10}; see Figure~\ref{fig:061201spec}.

\begin{figure}
\epsscale{1}
\plotone{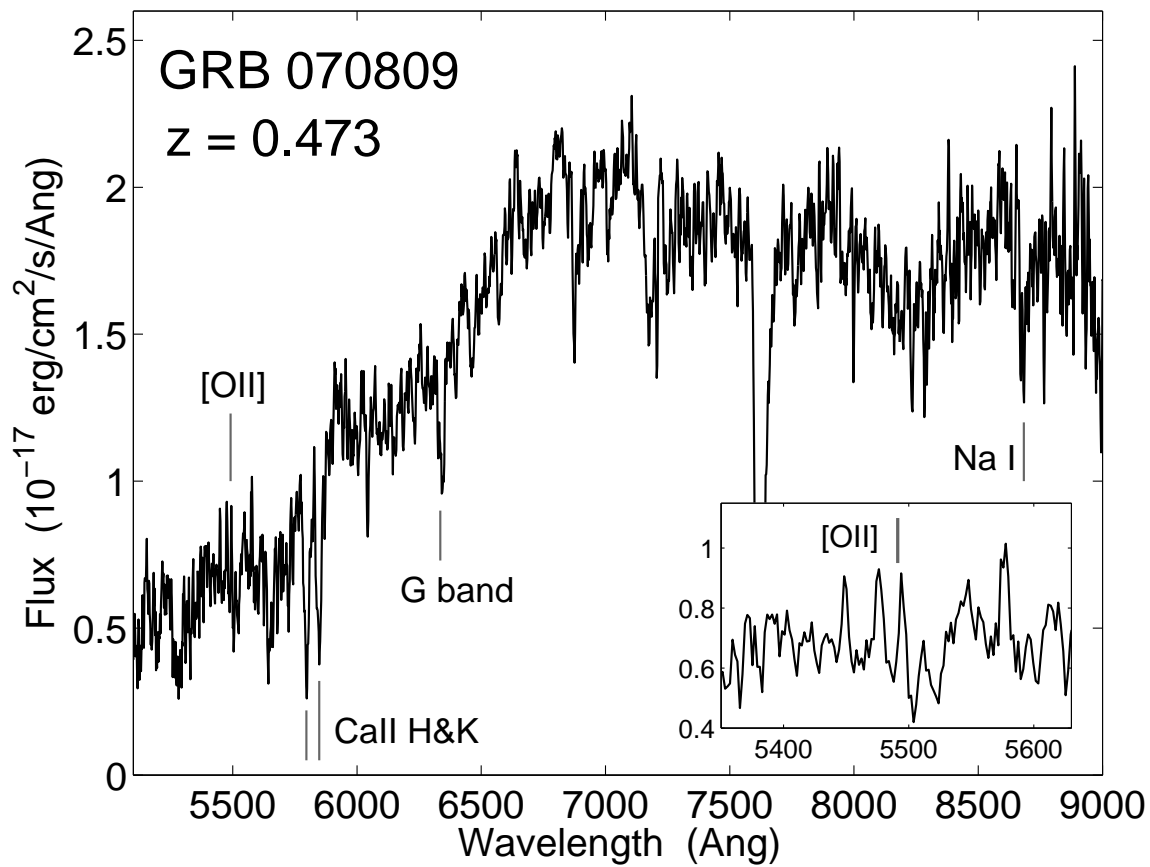}
\caption{Magellan/LDSS3 spectrum of the galaxy with the lowest
probability of chance coincidence near the position of GRB\,070809.
This galaxy is marked ``S3'' in Figure~\ref{fig:070809}.  It has a
redshift of $z=0.473$ and is an early-type galaxy with no evidence for
on-going star formation activity (see inset).
\label{fig:070809spec}} 
\end{figure}

For GRB\,070809 we obtained spectra of two galaxies located
$5.9\arcsec$ and $6.0\arcsec$ from the optical afterglow position
(marked ``S2'' and ``S3'', respectively in Figure~\ref{fig:070809})
using LDSS3 on 2008 January 14 UT.  The galaxy at a separation of
$5.9\arcsec$ was previously identified as a star forming galaxy at
$z=0.218$ by \citet{gcn7889}.  Here we find that the object at a
separation of $6.0\arcsec$ is an early-type galaxy at $z=0.473$, with
no evidence for on-going star formation activity
(Figure~\ref{fig:070809spec}).

\begin{figure}
\epsscale{1}
\plotone{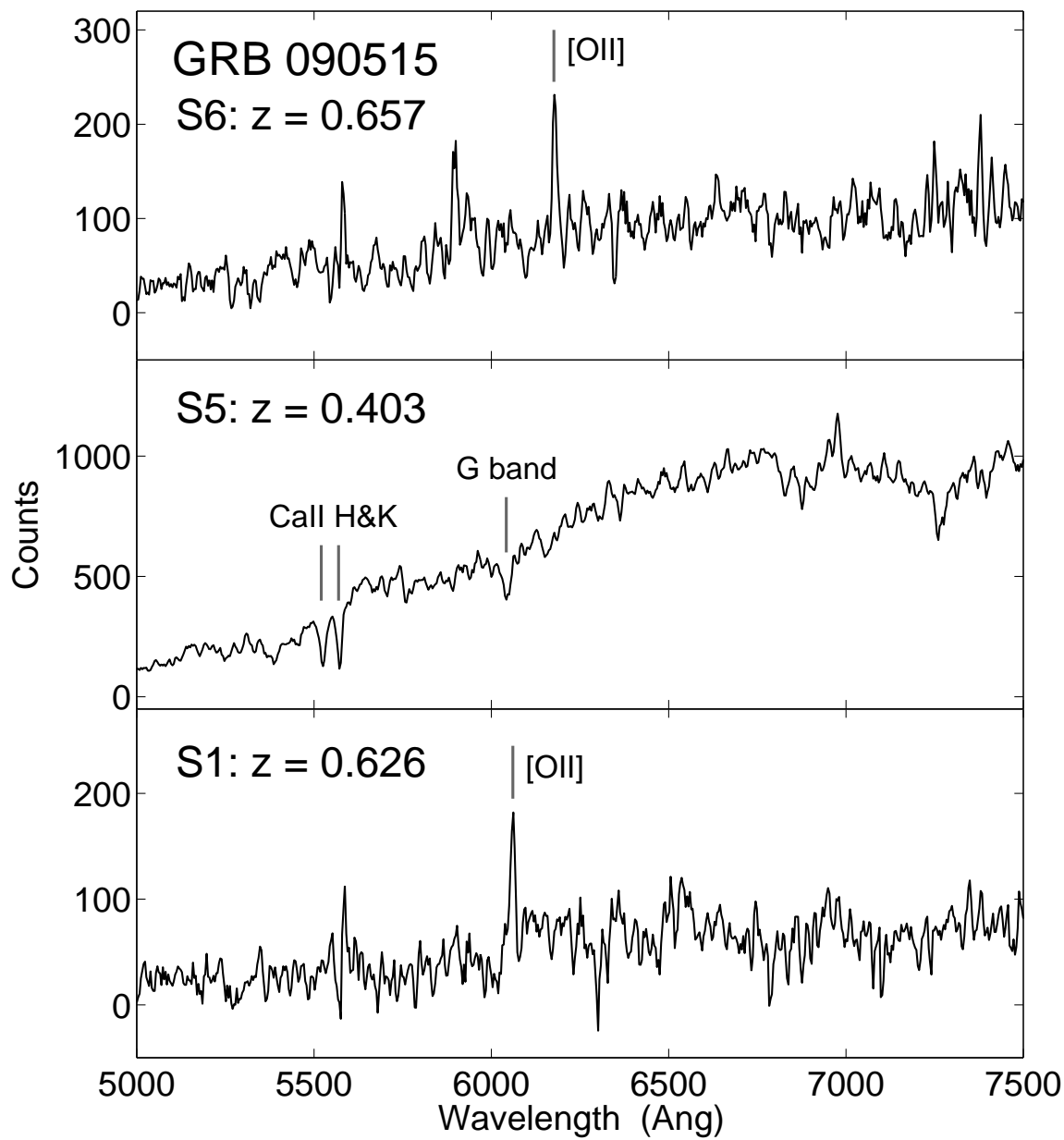}
\caption{Magellan/LDSS3 spectra of three galaxies with a low
probability of chance coincidence near the position of GRB\,090515.
The galaxy with the lowest probability of chance coincidence is marked
``S5'' in Figure~\ref{fig:090515}.  It has a redshift of $z=0.403$ and
is an early-type galaxy which is part of a galaxy cluster (Fong et
al. in prep.)
\label{fig:090515spec}} 
\end{figure}

Finally, for GRB\,090515 we obtained multi-object spectroscopic
observations with LDSS3 for nearly 100 galaxies within a
$5\arcmin\times 5\arcmin$ field centered on the GRB position.  These
observations provide redshifts for several galaxies near the host
position, including a star forming galaxy at $z=0.626$ ($5.8\arcsec$
offset; ``S1'' in Figure~\ref{fig:090515}), an early-type galaxy at
$z=0.403$ ($14.0\arcsec$ offset; ''S5''), and a star forming galaxy at
$z=0.657$ ($14.9\arcsec$ offset; ``S6''); see
Figure~\ref{fig:090515spec}.  In an upcoming paper we will demonstrate
that the galaxy at $z=0.403$ is a member of a cluster.

\section{Afterglow Properties}
\label{sec:ag}

We begin our investigation by assessing the distribution of optical
afterglow magnitudes for the three samples defined in \S\ref{sec:obs}.
The observed magnitudes and limits as a function of time after the
burst are shown in Figure~\ref{fig:optag1}.  The median observation
time for the sample is about 8.5 hr after the burst.  The distribution
of detected optical afterglow magnitudes is broad, ranging from
$r_{\rm AB}\approx 21$ to $\approx 26$ mag.  The limits range from
$r_{\rm AB}\gtrsim 23$ to $\gtrsim 25$ mag (with the exception of
GRB\,080702 which only has a shallow limit of about 21 AB mag:
\citealt{gcn7977}).

The mean brightness and standard deviation for {\it Sample 1}
are\footnotemark\footnotetext{These numbers remain essentially
unchanged if we extrapolate all measured magnitudes to the fiducial
time of 8.5 hr with a typical afterglow decay index of
$\alpha=-0.75$.}  $\langle r_{\rm AB}\rangle=23.0\pm 1.3$ mag.  This
is substantially fainter, by about an order of magnitude, than the
afterglows of long GRBs on a comparable timescale (e.g.,
\citealt{kkz+07,kkz+08}).  It is also remarkably similar to the
prediction of $R\approx 23$ mag at $t\approx 10$ hr by \citep{pkn01}.
Since the available limits are at least $\gtrsim 23$ mag, we conclude
that the bursts lacking optical detections are drawn from a population
with fainter afterglows.  For the 5 bursts in {\it Sample 2} we find a
median and standard deviation of $\langle r_{\rm AB}\rangle=24.4\pm
1.4$ mag, about 1.4 mag fainter than the bursts with coincident hosts.

The cumulative distributions for the three samples normalized to the
fiducial time of $8.5$ hr after the burst are shown in
Figure~\ref{fig:optag2}.  A Kolmogorov-Smirnov (K-S) test indicates
that there is only an $8\%$ probability that {\it Sample 1} and {\it
Sample 2} are drawn from the same underlying distribution of optical
afterglow brightnesses.  Similarly, the probability that {\it Sample
1} and {\it Sample 3} are drawn from the same underlying distribution
is $\lesssim 5\%$ (an upper limit since the bursts in {\it Sample 3}
are not detected in the optical).  On the other hand, the probability
that {\it Sample 2} and {\it Sample 3} are drawn from the same
underlying distribution is high, $\approx 50\%$.

The overall faintness of the optical afterglows in {\it Sample 2} and
{\it Sample 3} can be explained in two primary ways.  First, they
could result from systematically lower circumburst densities.  In the
standard afterglow model\footnotemark\footnotetext{We use the standard
synchrotron spectrum definitions: $\nu_m$ is the characteristic
synchrotron frequency corresponding to electrons with the minimum
Lorentz factor ($\gamma_m$) of the electron distribution, $N(\gamma)
\propto\gamma^{-p}$; $p$ usually has a value of $\approx 2.2-2.5$; and
$\nu_c$ is the synchrotron cooling frequency \citep{spn98}.} with
$\nu_m<\nu_{\rm opt}<\nu_c$, the afterglow flux scales as
$F_\nu\propto n^{1/2}$ for a uniform medium \citep{spn98}.  Thus, a
difference of about $+1.4$ mag can be explained with a density that is
about an order of magnitude lower than for the bursts in {\it Sample
1}.

Alternatively, the fainter fluxes may be due to higher redshifts for
{\it Sample 2} and {\it Sample 3} compared to {\it Sample 1} since the
optical flux also scales as $F_\nu\propto (1+z)^{(3+p)/4}\,d_L^{-2}$,
where $d_L$ is the luminosity distance.  The $+1.4$ mag difference
corresponds to $\Delta z\approx +0.5$ ($+1$) for a {\it Sample 1} mean
redshift of $z=0.5$ ($z=1$).  Similarly, the flux also depends on the
total energy, with $F_\nu\propto E^{(3+p)/4}$, and therefore a
difference of $+1.4$ mag can be explained with an energy release lower
by about a factor of 3.

The various scenarios (lower density, lower energy, or higher
redshift) can be further explored through a comparison of the prompt
$\gamma$-ray emission, and the relation between the optical and X-ray
afterglow brightness.  In the framework of the standard GRB model we
do not expect lower densities to impact the prompt emission since it
is expected to be produced by internal processes (shocks or magnetic
dissipation) that do not depend on the external medium.  On the other
hand, lower energies or higher redshifts will tend to systematically
affect the prompt and afterglow emission.

\begin{figure}
%\epsscale{1}
%\plotone{fig11.ps}
\centering
\includegraphics[angle=0,width=5.0in]{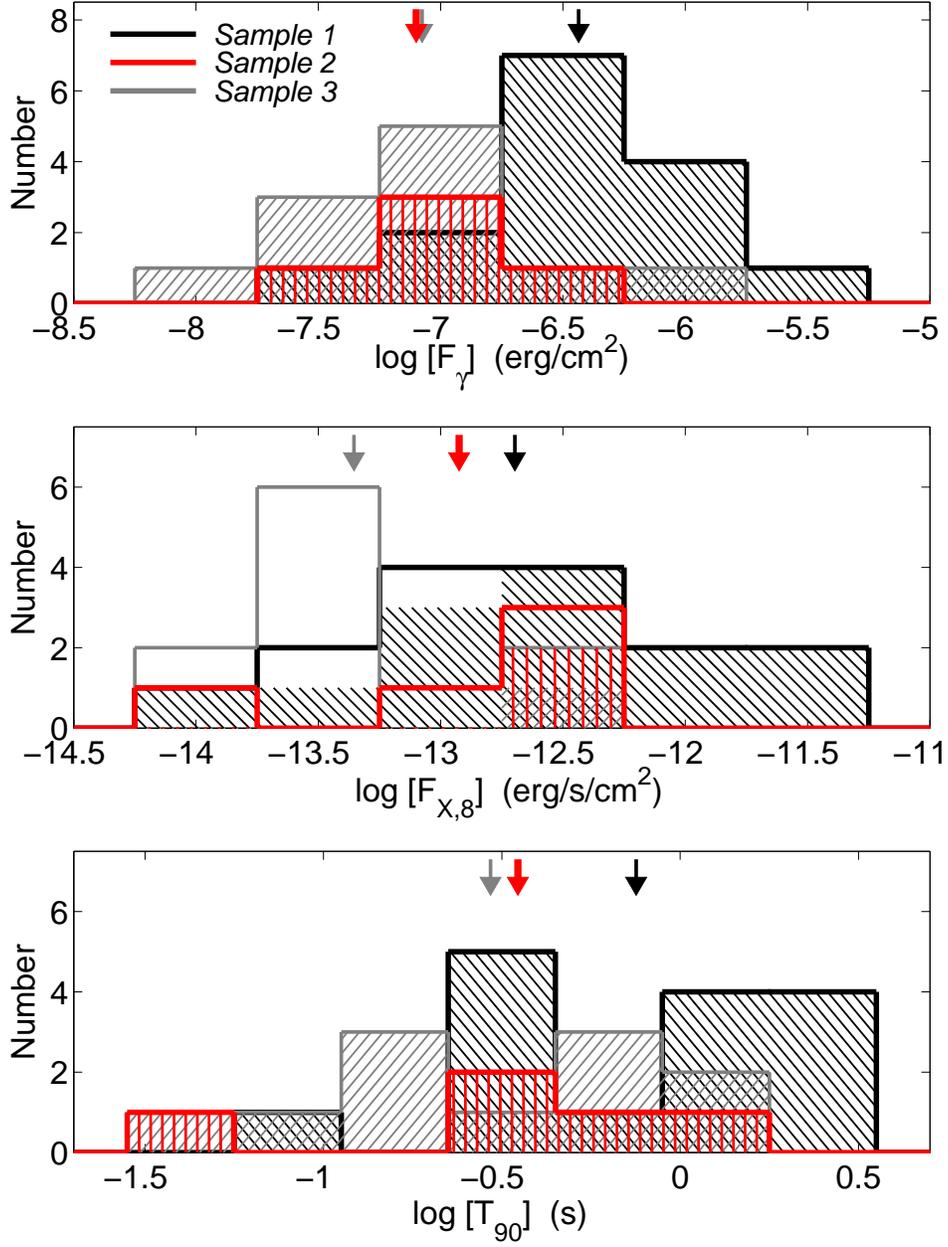}
\caption{Histograms of $\gamma$-ray fluence (top), afterglow X-ray
flux at 8 hr (middle), and duration (bottom) for the three short GRB
samples discussed in this paper.  The arrows mark the mean for each
sample, indicating that the bursts in {\it Sample 2} and {\it Sample
3} have lower $\gamma$-ray fluences, fainter X-ray fluxes, and shorter
durations, than the bursts with detected afterglows and coincident
hosts.
\label{fig:fg_fx_t90}} 
\end{figure}

In Figure~\ref{fig:fg_fx_t90} we plot the distributions of
$\gamma$-ray fluence ($F_\gamma$), afterglow X-ray flux at the
fiducial time of 8 hr ($F_{X,8}$), and duration ($T_{90}$) for the
three samples.  These distributions allow us to explore the underlying
reason for the difference in optical afterglow brightness between the
three samples.  First, we find that the distribution of $F_\gamma$
values for {\it Sample 1} has a mean value that is about a factor of 5
times larger than for {\it Sample 2} and {\it Sample 3}.  This is
indicative of higher redshifts for the latter two samples if the
isotropic-equivalent energies of all short GRBs are similar, or
alternatively a lower energy scale if the redshifts are similar.

In the same vein, we find that the distributions of $F_{X,8}$ values
for {\it Sample 2} and {\it Sample 3} have lower means than for {\it
Sample 1}.  This result is indicative of overall fainter afterglow
emission for the former two samples, and this can again be explained
in the context of lower energies or higher redshifts.  Unlike in the
case of the $\gamma$-ray emission, a lower circumburst density would
also account for the fainter X-ray fluxes if $\nu_X<\nu_c$.  Finally,
we find that the durations of the bursts in {\it Sample 2} and {\it
Sample 3} are shorter by about a factor of 2 compared to the events
with optical afterglows and coincident hosts, although there is
substantial scatter in all three samples.  The shorter durations are
not trivially explained in the context of lower energies, higher
redshifts, or lower densities.

\begin{figure}
\epsscale{1}
\plotone{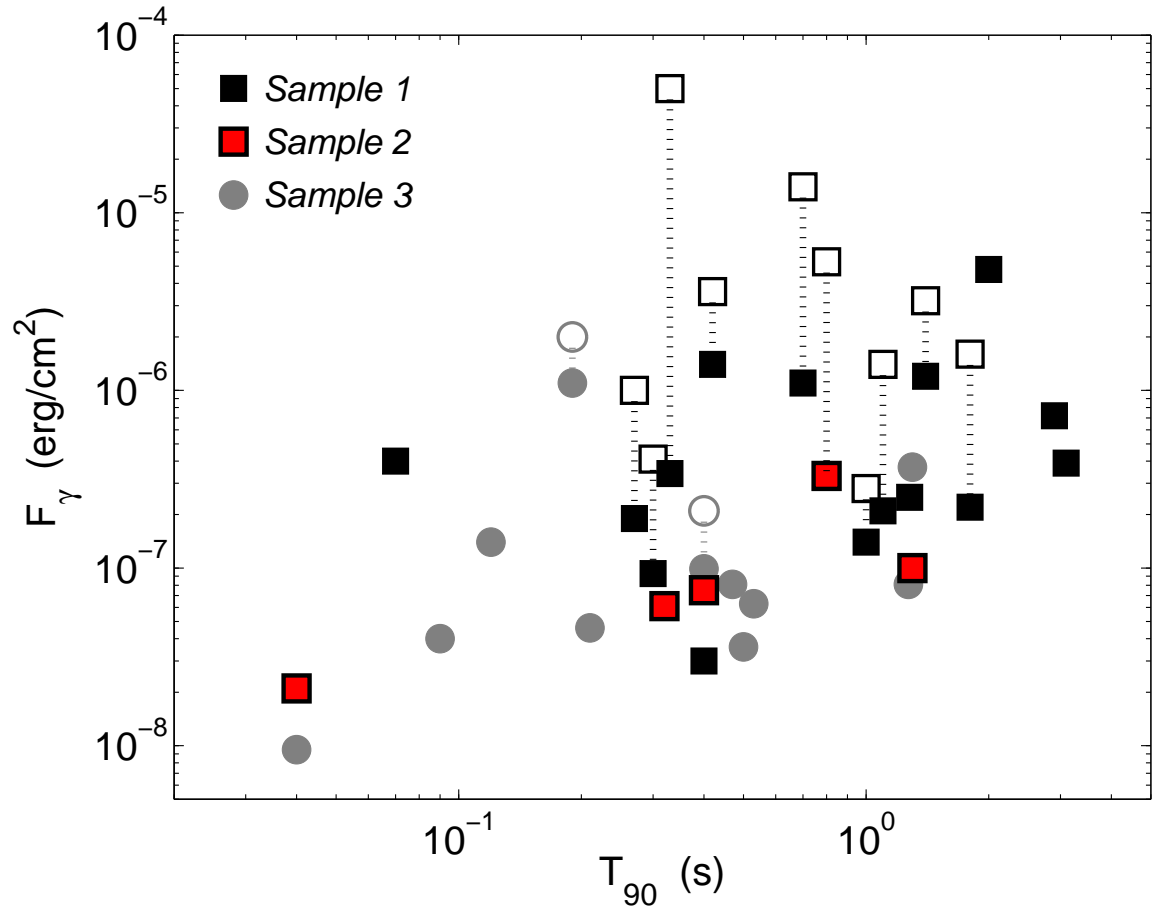}
\caption{Short GRB $\gamma$-ray fluence as a function of duration for
the three samples discussed in this paper.  An overall correlation is
apparent in the data.  The bursts in {\it Sample 2} and {\it Sample 3}
appear to lie below the mean correlation for the bursts in {\it Sample
1}, i.e., they have lower fluences for their durations, or longer
durations for their fluences.
\label{fig:t90_fg}} 
\end{figure}

To explore this result in more detail we plot the observed fluence as
a function of duration for the events in all three samples
(Figure~\ref{fig:t90_fg}).  We find that there is a mild positive
correlation between the two quantities, but that the events in {\it
Sample 1} appear to have larger fluences at a given duration compared
to the events in {\it Sample 2} and {\it Sample 3}.  This is
indicative of lower $\gamma$-ray fluxes for the latter two samples,
possibly as a result of higher redshifts.  Higher redshifts will also
shift the intrinsic durations of the bursts in {\it Sample 2} and {\it
Sample 3} into better agreement with the bursts in {\it Sample 1}.

We therefore conclude that the differences in prompt emission and
optical/X-ray afterglow properties are consistent with a higher
redshift origin for the bursts in {\it Sample 2} and {\it Sample 3}.
The fainter afterglow emission is also consistent with lower density
environments for these two samples, although this does not clearly
explain the differences in prompt emission (at least in the framework
of the standard GRB model).  We return to the discussion of low
density versus a high redshift origin in \S\ref{sec:hosts} when we
investigate the host galaxy properties.

In addition to the overall faintness of the optical and X-ray
afterglows, a substantial difference in density may also be imprinted
on the {\it ratio} of optical to X-ray brightness.  This is because
the synchrotron cooling frequency depends on density as $\nu_c\propto
n^{-1}$, and is therefore expected to transition across the X-ray band
as the density decreases.  For $\nu_c>\nu_X$ the X-ray and optical
bands occupy the same portion of the synchrotron spectrum, with a
resulting spectral index of $\beta=-(p-1)/2\approx -0.6$ to $\approx
-0.75$ while for $\nu_c<\nu_X$ (i.e., high density), the spectrum
between the two bands will be steeper, reaching a maximum value of
$\approx -1.25$ when $\nu_c\approx\nu_{\rm opt}$.  In
Figure~\ref{fig:opt_xray} we plot the X-ray flux versus optical
magnitude for all three samples.  For each burst the fluxes in the
optical and X-rays are taken at similar times after the burst, or
extrapolated to a common time.  The correction factors due these
extrapolations are generally\footnotemark\footnotetext{We do not
extrapolate X-ray upper limits.} $\lesssim 2$ (Table~\ref{tab:shbs}).
For the combined {\it Sample 1} and {\it Sample 2} we find a clear
correlation between the fluxes in the two bands, leading to a mean
optical to X-ray spectral index of $\langle\beta_{\rm
OX}\rangle=-0.72\pm 0.17$.  This is essentially indistinguishable from
the ratio for long GRBs, $\langle\beta_{\rm OX}\rangle=-0.65\pm 0.35$
\citep{jhf+04}.  The median values are consistent with $\nu_c\gtrsim
\nu_X$, but exhibit dispersion that is likely due to scatter in the
values of $p$ and/or the location of $\nu_c$ relative to the X-ray
band.

The similarity of $\beta_{\rm OX}$ for long and short GRBs does not
necessarily indicate that the densities are similar for the two
samples.  In particular, if $\nu_c$ is located close to the X-ray band
for long GRBs, while for short GRBs $\nu_c\gg\nu_{\rm X}$ (due to a
lower density), the effect on $\beta_{\rm OX}$ will be marginal,
particularly within the overall observed scatter.  For example, with
$p=2.5$ the difference in $\beta_{\rm OX}$ between a model with
$\nu_c$ exactly intermediate between the optical and X-ray bands, and
a model with $\nu_c>\nu_{\rm X}$, is $\Delta\beta_{\rm OX}\approx
0.25$.  On the other hand, the scatter resulting from a range of
$p=2.2-2.5$ is of the same order, $\Delta\beta_{\rm OX}=0.15$.
Similarly, the nearly equivalent median $\beta_{\rm OX}$ values may
indicate that for both GRB populations $\nu_c>\nu_{\rm X}$.  In this
case, the resulting lower limits on $\nu_c$ therefore prevent the use
of $\beta_{\rm OX}$ as an indicator of density.

Comparing {\it Sample 1} and {\it Sample 2}, we find no clear
difference in $\beta_{\rm OX}$ (Figure~\ref{fig:opt_xray}).  The same
is true for the bursts in {\it Sample 3}, which are all consistent
with the same relation given the optical upper limits and a mix of
X-ray detections and upper limits.  Thus, the ratio of optical to
X-ray flux does not allow us to distinguish redshift/energy and
density effects between the three samples.

To summarize, the optical afterglows of short GRB without coincident
hosts (or with only optical limits) are systematically fainter than
those of short GRBs with coincident hosts.  The same is true for their
X-ray fluxes and $\gamma$-ray fluences.  The fainter afterglows may
reflect lower densities (by an order of magnitude), but this does not
naturally explain the lower $\gamma$-ray fluences.  Alternatively, the
fainter afterglows and $\gamma$-ray fluences can be explained as a
result of higher redshifts ($\Delta z\approx 0.5-1$) or lower energies
(by about a factor of 3).

\section{Large-Scale Environments}
\label{sec:hosts}

We next turn to an analysis of the large-scale environments of the
bursts in {\it Sample 2}, partly in comparison to the hosts of bursts
in {\it Sample 1}.  As indicated in \S\ref{sec:obs}, we place limits
of $r_{\rm AB}\approx 25.4-26.5$ mag on the brightness of any galaxy
underlying the five short GRB positions (Table~\ref{tab:nohost}).
These limits are several magnitudes fainter than the measured
brightnesses of short GRB hosts at $z\lesssim 1$ ({\it Sample 1}).

\subsection{Probabilities of Chance Coincidence}
\label{sec:prob}

To assess the potential that galaxies near each of the 5 bursts are
the hosts, we calculate their probability of chance coincidence.  We
follow the methodology of \citet{bkd02}, namely, we determine the
expected number density of galaxies brighter than a measured
magnitude, $m$, using the results of deep optical galaxy surveys
\citep{hpm+97,bsk+06}:
\begin{equation}
\sigma(\le m)=\frac{1}{0.33\times {\rm ln}(10)}\times
10^{0.33(m-24)-2.44} \,\,\,\,{\rm arcsec}^{-2}.
\label{eqn:gal}
\end{equation}
The probability for a given separation, $P(<\delta R)$, is then given
by
\begin{equation}
P(<\delta R)=1-{\rm e}^{-\pi (\delta R)^2\sigma(\le m)},
\label{eqn:prob}
\end{equation}
where we use the fact that for offsets substantially larger than the
galaxy size, $\delta R$ is the appropriate radius in
Equation~\ref{eqn:prob} \citep{bkd02}.

\begin{figure}
\epsscale{1}
\plotone{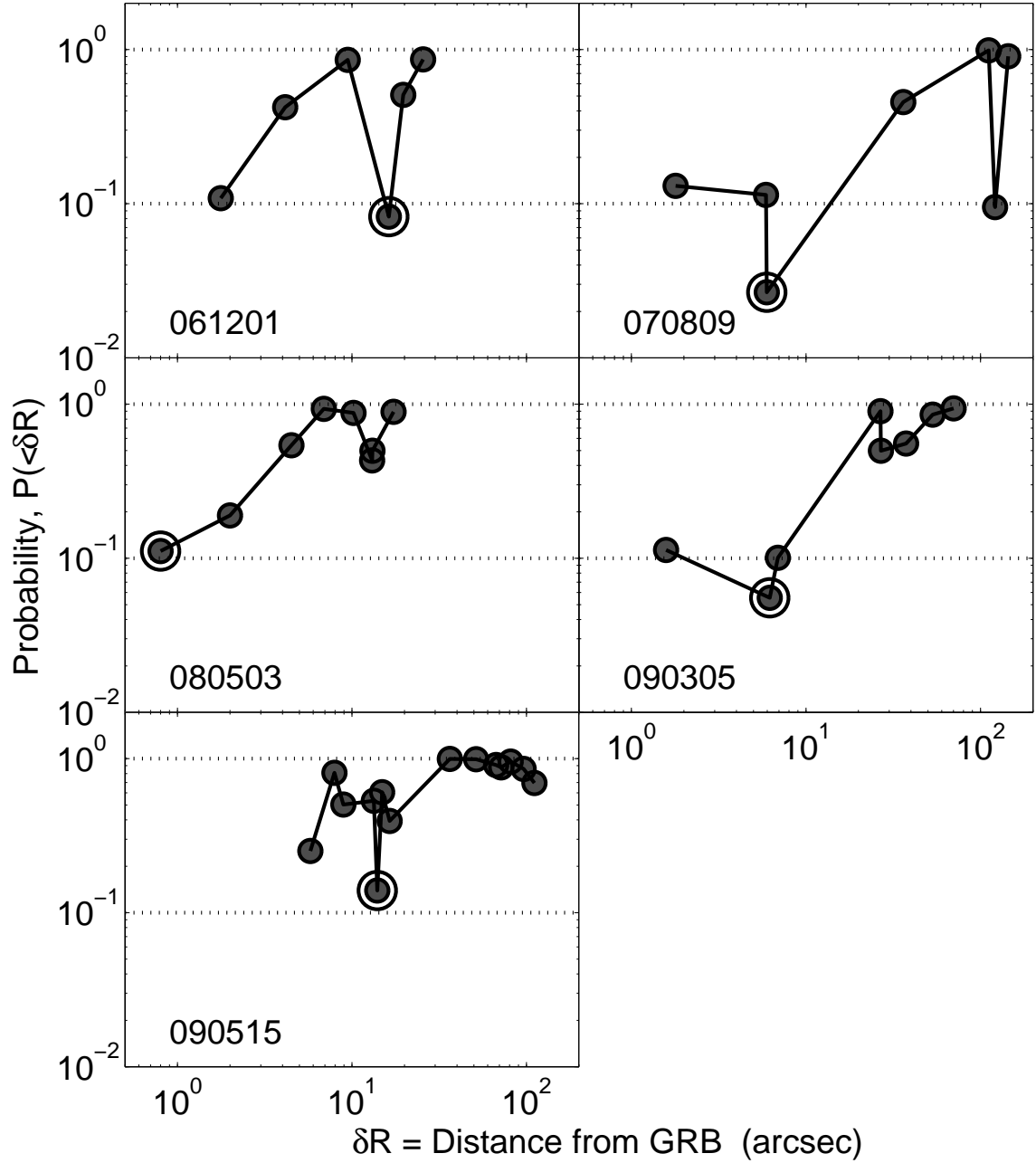}
\caption{Probability of chance coincidence as a function of distance
from a short GRB optical afterglow position for galaxies near the
location of each burst.  These are the galaxies marked in
Figures~\ref{fig:061201}-\ref{fig:090515}.  In each panel we mark the
galaxy with the lowest probability of chance detection with a circle.
In 4 of the 5 cases, the lowest probability is associated with
galaxies that are offset by $\sim 5-15\arcsec$.  Moreover, even the
nearest galaxies are offset by $\approx 1.6-5.8\arcsec$.
\label{fig:prob1}} 
\end{figure}

The resulting distributions for each field are shown in
Figure~\ref{fig:prob1}.  We include all galaxies that have
probabilities of $\lesssim 0.95$.  We find that for 4 of the 5 bursts,
faint galaxies ($\sim 25-26$ mag) can be identified within $\approx
1.6-2\arcsec$ of the afterglow positions, with associated chance
coincidence probabilities of $\approx 0.1-0.2$; in the case of
GRB\,090515 we do not detect any such faint galaxies within $\approx
5\arcsec$ of the afterglow position.  For GRB\,080503 we also include
the galaxy at an offset of $0.8\arcsec$ and $m_{\rm AB}({\rm
F606W})=27.3\pm 0.2$ mag identified by \citet{pmg+09} based on their
deeper stack of HST observations.  On the other hand, for 4 of the 5
bursts we find that the galaxies with the lowest probability of chance
coincidence, $\approx 0.03-0.15$, are brighter objects with offsets of
about $6-16\arcsec$ from the burst positions; only in the case of
GRB\,080503 the lowest chance coincidence is associated with the
nearest galaxy (see \citealt{pmg+09}).

\begin{figure}
\epsscale{1}
\plotone{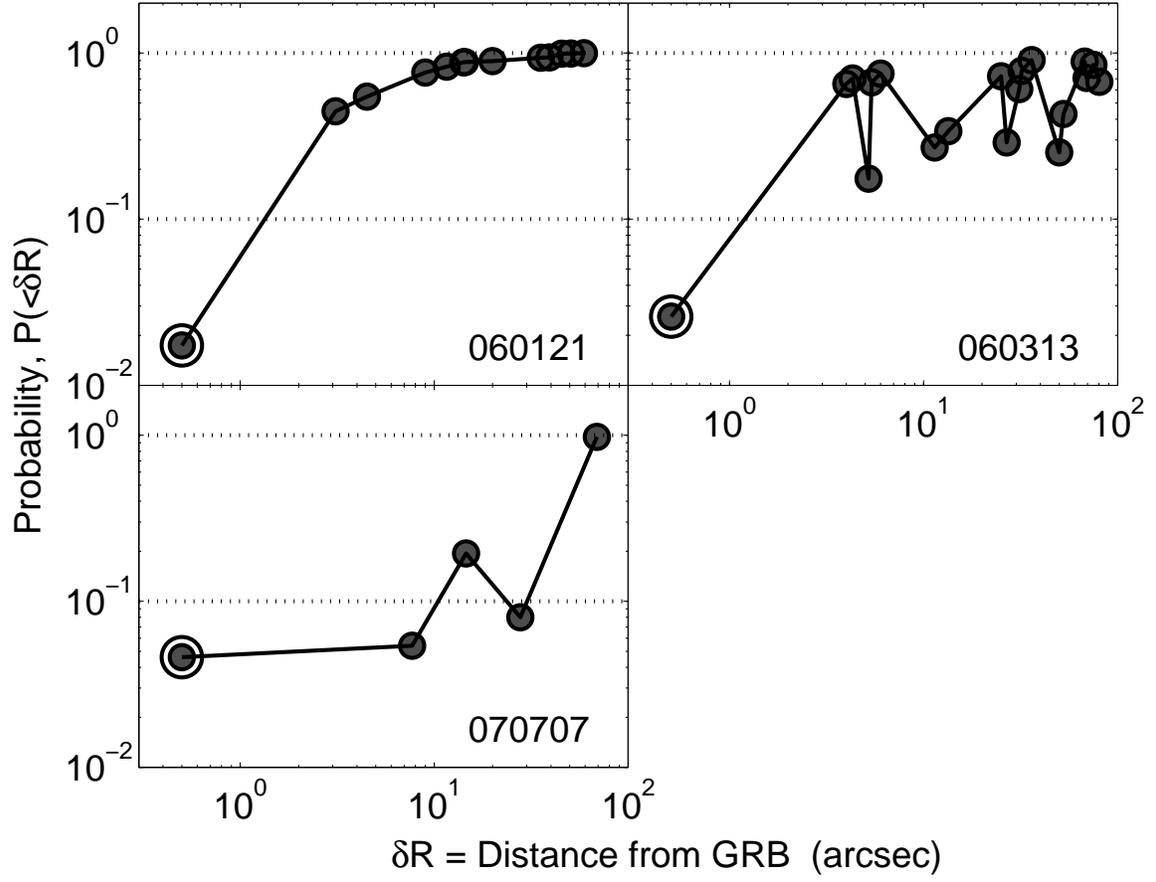}
\caption{Same as Figure~\ref{fig:prob1}, but for short GRBs with
coincident faint hosts.  In this case, the lowest probability of
chance coincidence is associated with the underlying faint host.
\label{fig:prob2}} 
\end{figure}

For comparison we repeat the same analysis for short GRBs from {\it
Sample 1} which have faint coincident hosts (GRB\,060121: 26.0 AB mag;
GRB\,060313: 26.6 AB mag; and GRB\,070707: 27.3 AB mag).  The results
of the probability analysis are shown in Figure~\ref{fig:prob2}.  We
find that in all three cases, the coincident hosts exhibit the lowest
probability of chance coincidence, $\approx 0.02-0.05$.  Only in the
case of GRB\,070707 do we find galaxies with $\delta R\gtrsim {\rm
few}$ arcsec that have $P(<\delta R)\lesssim 0.1$.  Thus, these three
bursts are consistent with negligible offsets from faint galaxies,
presumably at $z\gtrsim 1$.

The use of {\it a posteriori} probabilities to assign {\it unique}
galaxy associations is fraught with difficulties.  First, for a given
apparent brightness, galaxies located further away from the GRB
position, potentially due to larger kicks and/or longer merger
timescales in the NS-NS merger framework, have higher probabilities of
chance coincidence.  Since we have no {\it a priori} model-independent
knowledge of the range of possible kicks and merger timescales, we
cannot rule out galaxies at very large offsets for which $P(<\delta
R)\sim 1$.  Indeed, a reasonable constraint of $v_{\rm kick}\lesssim
10^3$ km s$^{-1}$ and $\tau_{\rm merger}\lesssim 10$ Gyr leads to only
a weak constraint on the offset of $\lesssim 10$ Mpc.  At $z=0.1$
($z=1$) this corresponds to about $1.5^\circ$ ($0.3^\circ$), a
projected distance at which nearly all galaxies will have a chance
coincidence probability of order unity.

A second difficulty is that we are using angular offsets, which ignore
the potential wide range of redshifts (and by extension also
luminosities) of the various galaxies.  For example, if the faint
galaxies with small offsets are located at $z\gtrsim 1$, the
corresponding physical offsets are $\sim 15$ kpc, while if the
galaxies at $\sim 10\arcsec$ offsets are located at $z\sim 0.3$ (see
\S\ref{sec:obs}), the offsets are only somewhat larger, $\sim 30$ kpc.
A galaxy at an even lower redshift, $z\sim 0.1$, with an offset of 50
kpc will be located about $30\arcsec$ from the GRB position and incur
a large penalty in terms of chance coincidence probability.  It is
important to note, however, that galaxies at lower redshift will
generally have brighter apparent magnitudes, partially compensating
for the larger angular separations (Equations~\ref{eqn:gal} and
\ref{eqn:prob}).  In only a single case (GRB\,070809) we find a galaxy
with $P(<\delta R)\lesssim 0.1$ at $\delta R\gtrsim 1'$ (which at
$z=0.043$ for this galaxy corresponds to a physical offset of about
100 kpc).

A final complication, which is not unique to this subset of events, is
that we can only measure projected offsets, $\delta R=\delta R_{\rm
3D}\times {\rm cos}(\theta)$.  The measured offsets can be used as
lower limits on the actual offsets, while for the overall distribution
we can apply an average correction factor of $\pi/2$, based on the
expectation value for the projection factor, ${\rm cos}(\theta)$.

Despite these caveats we can address the probability that {\it all} of
the associations are spurious.  This joint probability is simply the
product of the individual probabilities \citep{bkd02}.  For the faint
galaxies at small angular separations the probability that all are
spurious associations is $P_{\rm all}\approx 8\times 10^{-5}$, while
for the galaxies with the lowest probability of chance coincidence the
joint probability is nearly 30 times lower, $P_{\rm all}\approx
3\times 10^{-6}$.  Conversely, the probabilities that {\it none} of
the associations are spurious are $\approx 0.42$ and $\approx 0.59$,
respectively.  These values indicate the some spurious coincidences
may be present for {\it Sample 2}.  Indeed, the probabilities that 1,
2, or 3 associations are spurious are $[0.40,\,0.15,\,0.027]$ and
$[0.34,\,0.068,\,0.006]$, respectively.  These results indicate that
for the faint galaxies it is not unlikely that $2-3$ associations (out
of 5) are spurious, while for the brighter galaxies $1-2$ associations
may be spurious.  This analysis clearly demonstrates why a joint
statistical study is superior to case-by-case attempts to associate
short GRBs with galaxies at substantial offsets.

We therefore conclude that despite the weaknesses inherent to {\it a
posteriori} probabilities, we conclude that there is stronger
statistical support for an association of at least some of the 5
bursts in {\it Sample 2} with bright galaxies at separations of $\sim
10\arcsec$, than for an association with the faint galaxies at
separations of $\sim 2\arcsec$.  Clearly, we cannot rule out the
possibility that in reality the hosts are a mix of faint and bright
galaxies with a range of angular offsets of $\gtrsim 2\arcsec$.  We
note that if deeper observations eventually lead to the detection of
underlying galaxies ($\lesssim 0.5\arcsec$) at the level of $\approx
27$ mag, the associated chance coincidence probabilities will be
$\approx 0.05$ per object, and the joint probabilities will be only
slightly higher than for the bright galaxies with $\sim 10\arcsec$
offsets.  On the other hand, if we can achieve magnitude limits of
$\gtrsim 28$ mag on any coincident hosts, the resulting probabilities
of chance coincidence will be larger than for the offset bright
galaxies.  Thus, eliminating the possibility of underlying hosts at
the level of $\gtrsim 28$ mag is of the utmost importance.  So far,
only GRB\,080503 has been observed to such a depth with no detected
coincident host \citep{pmg+09}, but observations of the full sample
are required for a robust statistical comparison.  As we discuss in
\S\ref{sec:highz} below, such deep limits will also reduce the
probability of underlying hosts based on redshift arguments.

Beyond the use of projected angular offsets, we note that the faint
galaxies near the burst positions are likely to have projected
physical offsets of about\footnotemark\footnotetext{Although the
redshifts of these galaxies are not known, the angular diameter
distance is nearly independent of redshift beyond $z\sim 1$, which is
appropriate for these faint host galaxies.} $15$ kpc.  To assess the
projected physical offsets for the galaxies with the lowest
probability of chance coincidence we measured spectroscopic redshifts
in three cases (GRBs 061201, 070809, and 090515; \S\ref{sec:obs}).  In
the case of GRB\,061201, the galaxy redshift of $z=0.111$ leads to a
projected physical offset of $32.4$ kpc \citep{fbf10}.  In the case of
GRB\,070809, the lowest probability of chance coincidence is not
associated with the star forming galaxy\footnotemark\footnotetext{We
note that even if the burst was associated with this galaxy, the
corresponding offset would be $20.6$ kpc.} at $z=0.218$ identified by
\citet{gcn7889}, but instead belongs to the early-type galaxy
identified here, which has a redshift of $z=0.473$, and hence an
offset of $34.8$ kpc.  Finally, for GRB\,090515, the lowest
probability of chance coincidence is associated with a cluster
early-type galaxy at $z=0.403$, leading to a physical offset of $75$
kpc.

Thus, two scenarios emerge from our investigation of the large-scale
environments: (i) the bursts are spatially coincident with
currently-undetected galaxies (with $r_{\rm AB}\gtrsim 26$ mag); or
(ii) the bursts have substantial offsets of at least $\approx 15-75$
kpc depending on whether they are associated with faint galaxies at
small angular separations, or brighter galaxies at $z\sim 0.1-0.5$;
for the ensemble of 5 events the larger offsets are statistically more
likely than the $\sim 15$ kpc offsets.  In the context of large
offsets, even larger values may be possible if the bursts originated
in galaxies with larger separations and $P(<\delta R)\sim 1$.

These two scenarios echo the possibilities that emerged from the
analysis of the afterglow and prompt emission properties
(\S\ref{sec:ag}).  Distinguishing between these possibilities is
clearly of fundamental importance to our understanding of short GRBs:
the former scenario will point to a population of very faint hosts
(likely at high redshifts), while the latter scenario will provide
evidence for large offsets (due to kicks or a globular cluster origin)
and hence NS-NS/NS-BH progenitors for at least some short GRBs.

\subsection{Scenario 1: Undetected Faint Hosts at High Redshift}
\label{sec:highz}

We can place upper limits on the redshifts of the GRBs in {\it Sample
2} based on their detections in the optical (i.e., the lack of
complete suppression by the Ly$\alpha$ forest).  The afterglow of
GRB\,061201 was detected in the ultraviolet by the {\it Swift}/UVOT
and it is therefore located at $z\lesssim 1.7$ \citep{rvp+06}.  The
remaining four bursts were detected in the optical $g$- or $r$-band,
and can therefore be placed at $z\lesssim 4$ (Table~\ref{tab:shbs}).

\begin{figure}
\epsscale{1}
\plotone{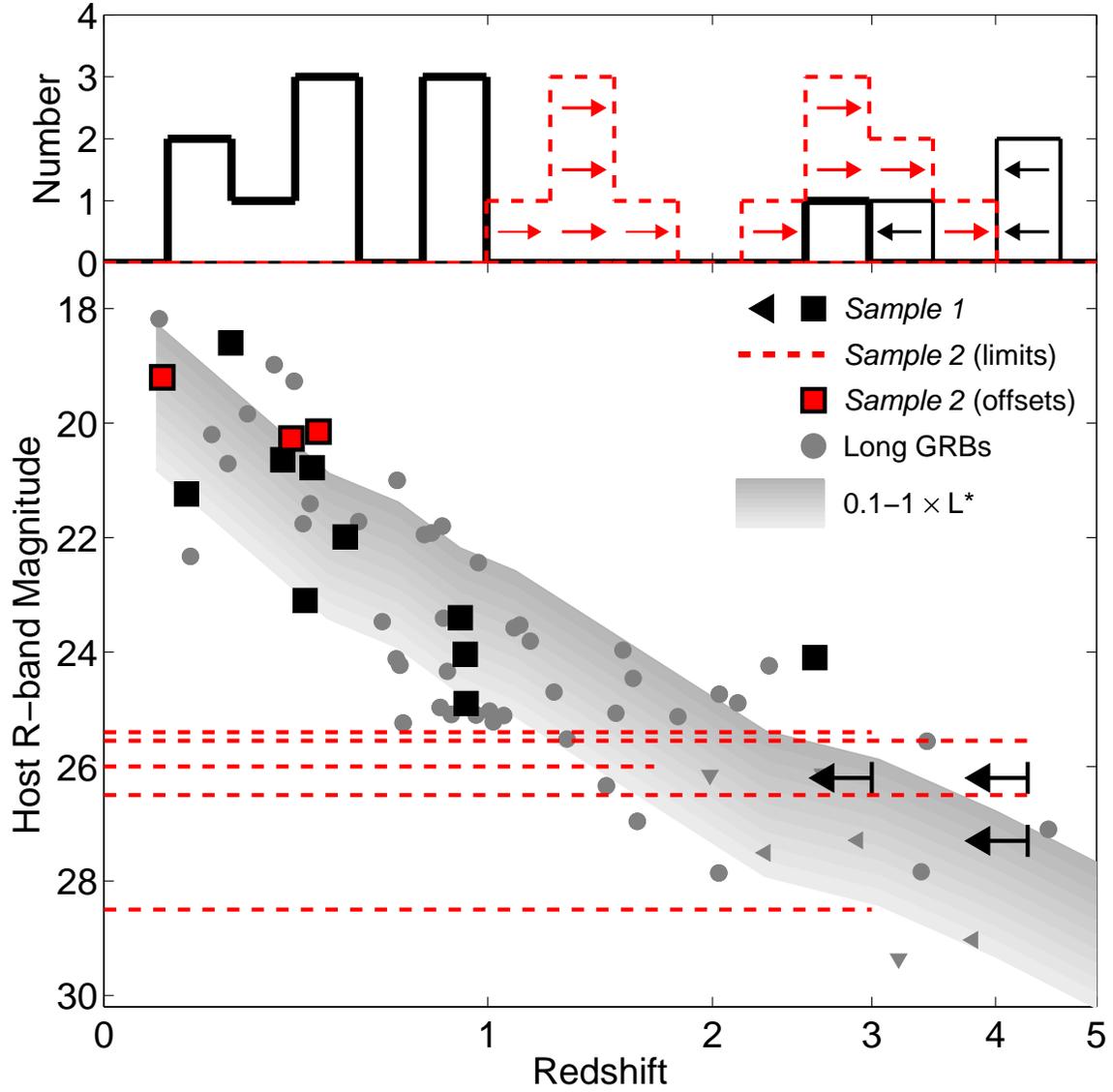}
\caption{Host galaxy optical magnitude as a function of redshift for
short GRB hosts (black squares), long GRB hosts (gray circles), and
galaxies with a luminosity of $0.1-1$ L$^*$ (shaded region).  The
dashed lines mark the upper limits at the GRB positions for the short
GRBs with no coincident hosts.  The arrows mark the upper limits on
the redshifts of three bursts with faint hosts, based on the detection
of the afterglows in the optical band (i.e., lack of a strong Lyman
break).  If underlying host galaxies exist for {\it Sample 2}, their
non-detection indicates $z\gtrsim 1.5$ (for $0.1$ L$^*$) or $\gtrsim
3$ (for L$^*$).  The alternative possibility that they are located at
similar redshifts to the detected hosts, requires $\lesssim 0.01$
L$^*$, but this does not naturally explain their fainter afterglows.
\label{fig:mags}} 
\end{figure}

We place additional constraints on the redshifts of any underlying
hosts using the existing sample short GRB host galaxies.  In
Figure~\ref{fig:mags} we plot the $r$-band magnitudes as a function of
redshift for the all available short GRB hosts from {\it Sample 1}.
For the faint hosts without known redshifts (GRBs 060121, 060313, and
070707), we place upper limits on the redshift using optical
detections of the afterglows (Table~\ref{tab:shbs}).  A wide range of
host magnitudes, $r_{\rm AB}\sim 16.5-27.5$ mag, is apparent.  We also
plot the $r$-band magnitudes of long GRB hosts \citep{sgl09}, as well
as the $r-z$ phase space that is traced by galaxies with luminosities
of $L=0.1-1$ L$^*$.  We use the appropriate value of L$^*$ as a
function of redshift, taking into account the evolving galaxy
luminosity function \citep{sag+99,bhb+03,wfk+06,rs09}.  We find
excellent correspondence between the hosts of long and short GRBs, and
the phase-space traced by $0.1-1$ L$^*$ galaxies, at least to $z\sim
4$.  In the context of these distributions, the available limits for
the short GRBs in {\it Sample 2} translate to redshifts of $z\gtrsim
1.5$ if they are 0.1 L$^*$ galaxies, or $z\gtrsim 3$ if they are L$^*$
galaxies.  The latter lower limits are comparable to the redshift
upper limits inferred from the afterglow detections.  We note that for
GRB\,080503, the limit of $\gtrsim 28.5$ mag \citep{pmg+09} places
even more stringent limits on the redshift of an underlying $0.1-1$
L$^*$ galaxy.

The redshifts of $z\gtrsim 1.5$ for putative 0.1 L$^*$ hosts are
consistent with the faintness of the optical afterglows, from which we
inferred $\Delta z\approx 0.5-1$ compared to {\it Sample 1}
(\S\ref{sec:ag}).  We note, however, that the one known short GRB at
$z\gtrsim 2$ (GRB\,090426; \citealt{lbb+10}) has a host galaxy
luminosity of $\sim 2$ L$^*$, which may suggest that the appropriate
redshift lower limits are $z\gtrsim 3$.

The possibility that the five bursts originated at $z\gtrsim 3$ leads
to a bimodal redshift distribution (Figure~\ref{fig:mags}).  Nearly
all of the bursts in {\it Sample 1} with a known redshift (9/10) have
$z\approx 0.2-1$, with a median of $\langle z\rangle\approx 0.5$; the
sole exception is GRB\,090426 at $z=2.61$.  The three bursts with
faint coincident hosts have upper limits of $z\lesssim 4$ from
afterglow detections, while lower limit of $z\gtrsim 1.5-2$ can be
placed on these hosts if they have $L\gtrsim 0.1$ L$^*$.  Adding the
{\it Sample 2} bursts with the assumption that they have $z\gtrsim 3$
will furthermore result in a population of short GRBs with a median of
$z\sim 3$, and leave a substantial gap at $z\sim 1-2$
(Figure~\ref{fig:mags}).  If the 5 bursts are instead hosted by $0.1$
L$^*$ galaxies, the inferred lower limits on the redshifts ($z\gtrsim
1.5$) lead to a potentially more uniform redshift distribution.

It is difficult to explain a bimodal redshift distribution with a
single progenitor population such as NS-NS binaries, without appealing
to, for example, a bimodal distribution of merger timescales.  Another
possibility is two distinct progenitor populations, producing bursts
of similar observed properties but with distinct redshift ranges.
While these possibilities are difficult to exclude, they do not
provide a natural explanation for the short GRB population.

A final alternative explanation is that any underlying hosts reside at
similar redshifts to the known hosts in {\it Sample 1} ($z\sim 0.5$),
but have significantly lower luminosities of $\lesssim 0.01$ L$^*$.
This scenario would not naturally explain why the bursts in {\it
Sample 2} have fainter optical and X-ray afterglows, as well as lower
$\gamma$-ray fluences.  We therefore do not consider this possibility
to be the likely explanation.

\subsection{Scenario 2: Large Offsets}
\label{sec:kick}

While higher redshifts may explain the lack of detected hosts, the
fainter afterglows, and the weaker $\gamma$-ray fluences of the bursts
in {\it Sample 2}, this scenario suffers from several difficulties
outlined above.  The alternative explanation is that the bursts
occurred at significant offsets relative to their hosts, and hence in
lower density environments that would explain the faint afterglow
emission (though possibly not the lower $\gamma$-ray fluences).  As we
demonstrated in \S\ref{sec:prob}, the offsets may be $\sim 2\arcsec$
($\sim 15$ kpc) if the bursts originated in the faint galaxies at the
smallest angular separations, or $\sim 10\arcsec$ ($\sim 30-75$ kpc)
if they originated in the brighter galaxies with the lowest
probability of chance coincidence.  Below we address the implications
of these two possible offset groups through a comparison to the
offsets measured for the bursts in {\it Sample 1} (e.g.,
\citealt{fbf10}).

\begin{figure}
\epsscale{1}
\plotone{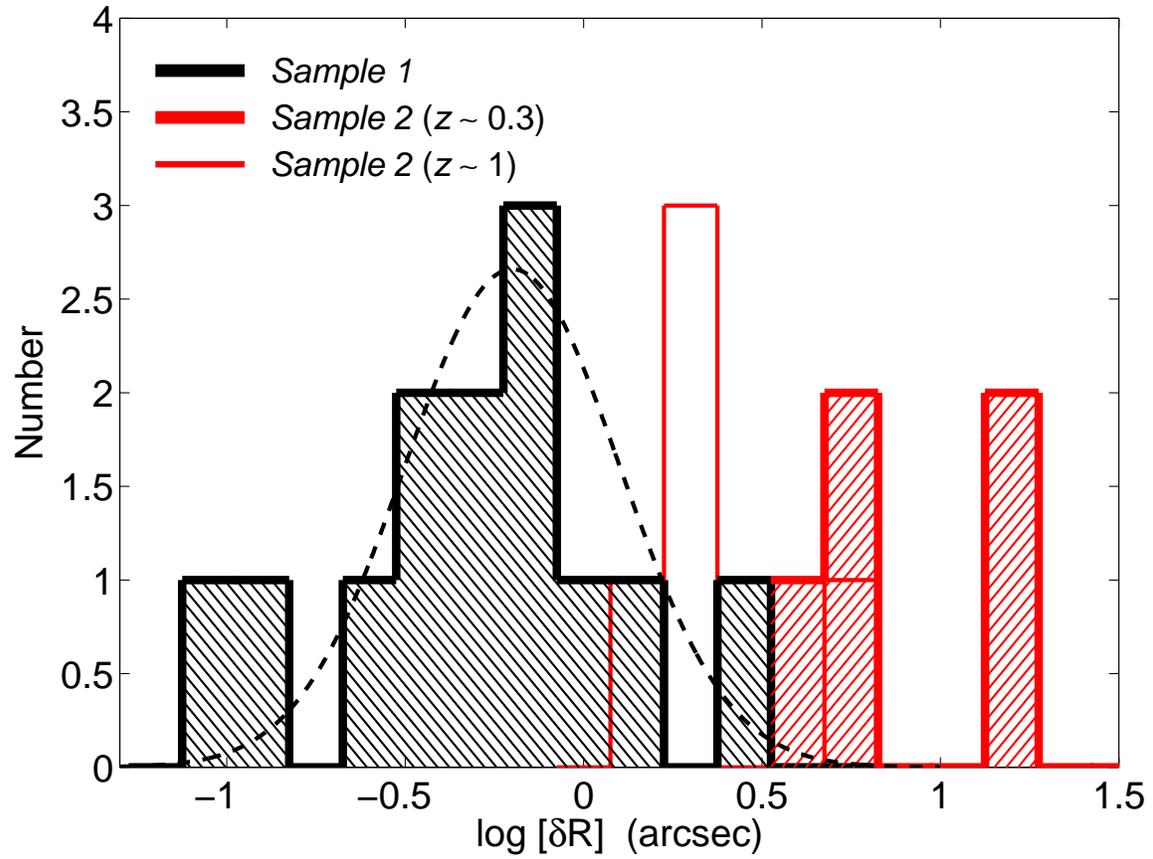}
\caption{Histogram of projected angular offsets relative to the host
galaxy center for short GRBs with coincident hosts (hatched black),
and bursts with no coincident hosts if the galaxies with lowest chance
coincidence probability are the hosts (hatched red), or if the faint
galaxies with smallest angular separation are hosts (open red); see
Figure~\ref{fig:prob1}.  The dashed line is a log-normal fit to the
bursts with coincident hosts.
\label{fig:offset1}} 
\end{figure}

We plot the distributions of projected angular offsets for the short
GRBs with and without coincident hosts in Figure~\ref{fig:offset1}.
The offsets for {\it Sample 1} have a mean and standard deviation of
about $0.7\pm 0.7\arcsec$ and a range of about $0.1-3\arcsec$.
Modeled with a log-normal distribution, the resulting mean and width
in units of arcseconds are ${\rm log}(\delta R)\approx -0.2$ and
$\sigma_{{\rm log}(\delta R)}\approx 0.35$.  If we associate the short
bursts in {\it Sample 2} with the galaxies that have the lowest
probability of chance coincidence, the resulting distribution has
$\langle\delta R\rangle=9\pm 6$ arcsec.  This is clearly distinct from
the distribution for {\it Sample 1}.  The effect is less pronounced if
we associate the bursts with the galaxies located at the smallest
angular offsets (Figure~\ref{fig:offset1}).  Even for these
separations the mean and standard deviation are ${\rm log}(\delta
R)\approx 2.6\pm 1.8$ arcsec.

\begin{figure}
\epsscale{1}
\plotone{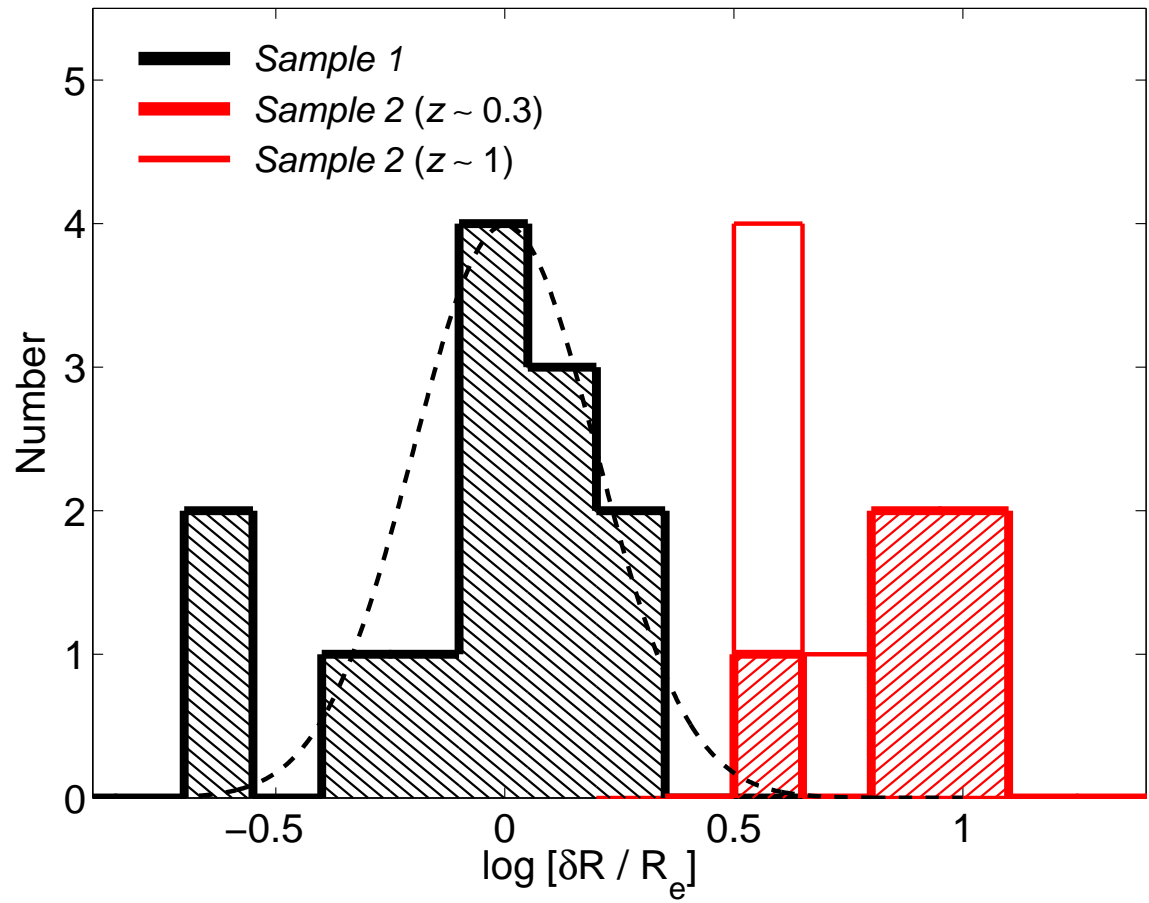}
\caption{Same as Figure~\ref{fig:offset1} but normalized relative to
the host effective radii, $R_e$.  The dashed line is a log-normal fit
to the bursts with coincident hosts.
\label{fig:offset2}} 
\end{figure}

As noted in \S\ref{sec:prob}, the projected angular distances may not
be the most robust quantity for measuring the offset distribution.  An
alternative quantity is the offset normalized by each host's effective
radius \citep{fbf10}.  This quantity takes into account the varying
sizes of the hosts due to both intrinsic size variations and redshift
effects.  It also gives a better indication of whether the burst
coincides with the host light or is significantly offset.  As shown in
Figure~\ref{fig:offset2}, the host-normalized offsets of {\it Sample
1} have a mean and standard deviation of about $1\pm 0.6$ $R_e$, and a
range of about $0.2-2$ $R_e$.  A log-normal fit results in a mean of
${\rm log}(\delta R/R_e)\approx 0$ and a width of $\sigma_{{\rm
log}(\delta R/R_e)}\approx 0.2$.  The bursts in {\it Sample 2} have
much larger host-normalized offsets, with $(\delta R/R_e)=7.3\pm 2.3$
if they originated in the galaxies with the lowest chance coincidence
probability.  Even if we associate the bursts with the nearest faint
hosts, the distribution has a mean of about $4$ $R_e$, reflecting the
fact that the effective radii of the faint galaxies are smaller than
those of the brighter ones.

\begin{figure}
\epsscale{1}
\plotone{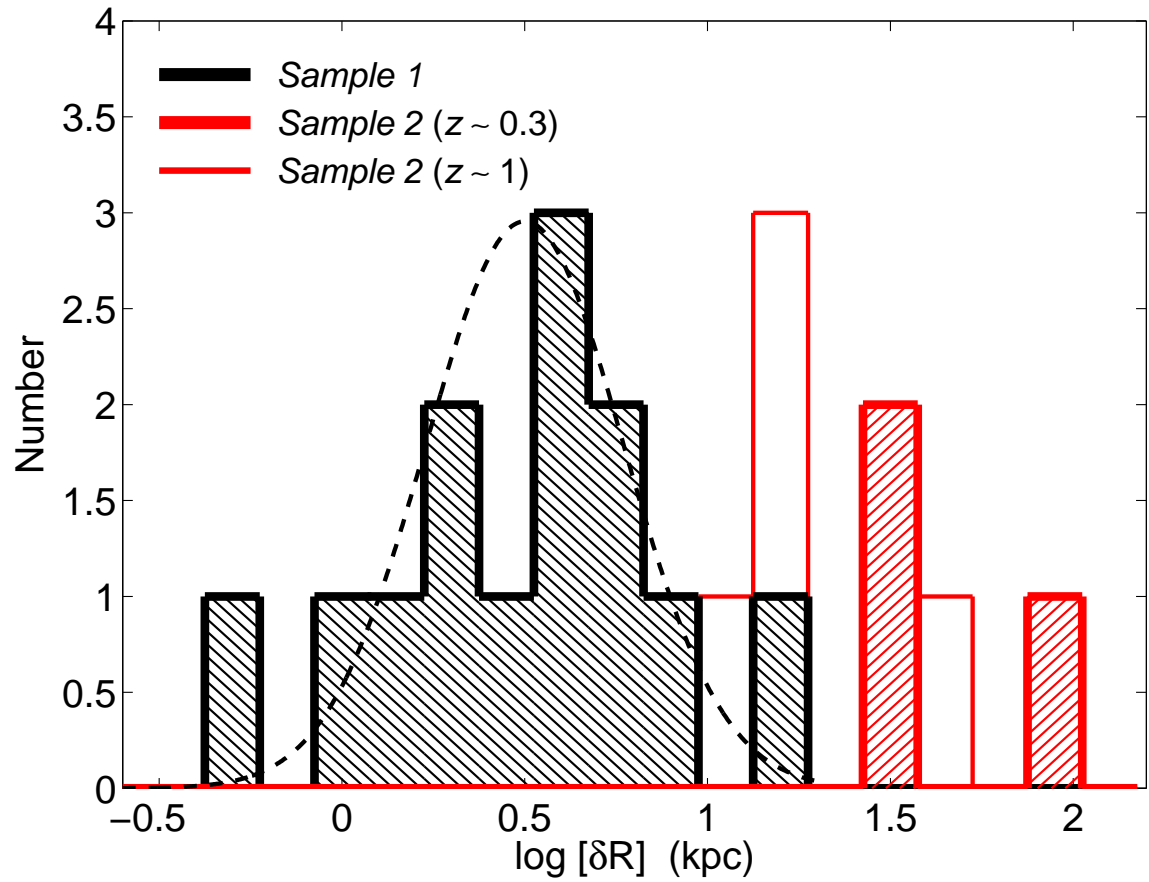}
\caption{Same as Figure~\ref{fig:offset1} but projected physical
offsets in units of kpc.  The dashed line is a log-normal fit to the
bursts with coincident hosts.
\label{fig:offset3}} 
\end{figure}

Finally, we plot the projected physical offsets in
Figure~\ref{fig:offset3}.  The mean and standard deviation for {\it
Sample 1} are $\delta R=4.2\pm 3.8$ kpc, and a log-normal fit results
in a mean of ${\rm log}(\delta R)\approx 0.5$ and a width of
$\sigma_{{\rm log}(\delta R)}\approx 0.3$.  On the other hand, the
bursts in {\it Sample 2} have a mean offset of about 19 kpc if they
arise in the faint galaxies with small angular separation, or about 40
kpc if they arise in the brighter galaxies, again pointing to distinct
distributions.

The distributions of angular, physical, and host-normalized offsets
exhibit a clear bimodality if we associate the bursts in {\it Sample
2} with the galaxies at $z\sim 0.1-0.5$.  This is particularly
apparent in the more meaningful quantities, namely physical and
host-normalized offsets (Figures~\ref{fig:offset2} and
\ref{fig:offset3}).  The effect is still apparent, though less
pronounced in the case of association with the faint galaxies at
$z\gtrsim 1$.  Thus, if the offset scenario is correct, the resulting
distributions point to a possible bimodality rather than a single
continuous distribution of offsets.  

The cumulative distributions of physical offsets for {\it Sample 1}
alone, and in conjunction with the two possible offset groups for {\it
Sample 2} are shown in Figure~\ref{fig:offset4}.  The combined
distributions have a median of about 4 kpc, driven by the bursts with
coincident hosts.  However, there is a clear extension to larger
physical offsets in the case of association with the brighter
galaxies, with about $20\%$ of all objects having $\delta R\gtrsim 30$
kpc.  The cumulative distributions are particularly useful for
comparison with NS-NS merger models since predictions exist for both
the kick scenario and the globular cluster origin model.  We turn to
this discussion below.

\section{Discussion and Implications}
\label{sec:disc}

We have shown that the lack of host galaxy detections in coincidence
with the bursts in {\it Sample 2} indicates either a high redshift
($z\gtrsim 2$) origin, or substantial projected offsets of $\sim
15-75$ kpc ($\approx 4-10$ $R_e$).  Both scenarios can account for the
lack of underlying galaxy detections and the fainter optical/X-ray
afterglows (either due to redshift or density effect); the possibility
of large offsets from galaxies at $z\sim 0.1-0.5$, however, does not
naturally explain the lower $\gamma$-ray fluences and somewhat shorter
durations of the bursts in {\it Sample 2}.  Indeed, the lower fluences
can be more easily accommodated in the case of undetected high
redshift coincident hosts, or in the case of faint hosts with offsets
of $\sim 15$ kpc.

Below we test the results of the offset scenario against predictions
for NS-NS merger models and dynamically-formed NS-NS binaries in
globular clusters.  We also investigate whether the densities expected
at offsets of $\sim 15-75$ kpc (i.e., galaxy halos or the
intergalactic medium) can accommodate the observed optical magnitudes.
Finally, we discuss the implications of the various scenarios for
short GRB energetics.

\subsection{Comparison to NS-NS Merger Models}

The sample of short GRBs with optical afterglows represents about
$1/3$ of all short bursts, and may thus not be fully representative.
One often-discussed bias is that the bursts with optical afterglows
require a high circumburst density, and therefore have negligible
offsets.  However, from our analysis in this paper it is clear that
one explanation for the lack of coincident hosts for the bursts in
{\it Sample 2} is indeed large offsets, {\it despite their detection
in the optical band}.

The bursts with only X-ray afterglow detections ({\it Swift}/XRT) have
typical positional uncertainties of $\sigma_X\approx 2-5\arcsec$ and
therefore lead to a deeper ambiguity about the identity of the hosts.
We have previously argued that most of these bursts are associated
with galaxies consistent with a negligible offset (or as large as tens
of kpc; \citealt{bfp+07,fbf10}).  In some cases larger offsets have
been advocated based on {\it a posteriori} chance coincidence
probabilities (e.g., GRB\,050509b: \citealt{bpp+06}; GRB\,060502b:
\citealt{bpc+07}).  However, from the analysis in \S\ref{sec:prob} it
is clear that while the inferred probabilities are only weakly
dependent on the positional accuracy when $\delta R\gtrsim\sigma_X$
(i.e., galaxies at large offsets), a large penalty is incurred for
faint galaxies located {\it within} the X-ray error circles since in
that case the appropriate value in Equation~\ref{eqn:prob} is $\delta
R=\sigma_X$.  As a result, the faint galaxies will have chance
coincidence probabilities of $P(<\delta R)\sim 1$.  To avoid this
complication in our comparison to NS-NS model predictions we restrict
the analysis to the sample with optical afterglow positions.

\begin{figure}
\epsscale{1}
\plotone{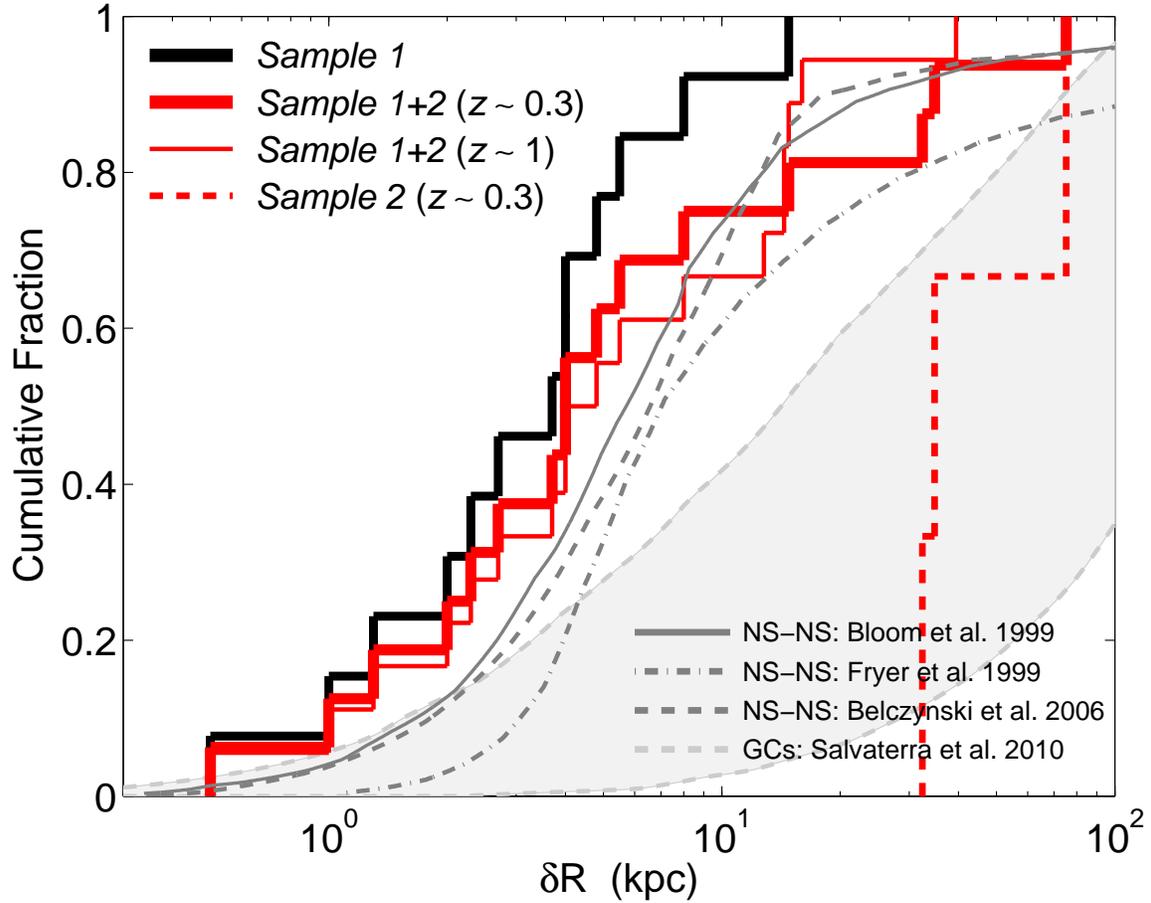}
\caption{Cumulative distributions of projected physical offsets for
short GRBs with coincident hosts (black line), and combined with
offsets for the hosts with the lowest probabilities of chance
coincidence (thick red line) or the faint hosts with smallest angular
offsets (thin red line).  Also shown are predicted distributions for
NS-NS kicks from several models \citep{bsp99,fwh99,bpb+06}, and for
dynamically-formed NS-NS binaries from globular clusters (shaded
region marks a range of predictions for host galaxy masses of $5\times
10^{10}-10^{12}$ M$_\odot$; \citealt{sdc+10}).  The models with kick
velocities are in good agreement with the measured offset distribution
for either set of galaxy associations, while the globular clusters
model provides a poor match to the data.
\label{fig:offset4}} 
\end{figure}

In recent work by \citet{fbf10} we have shown that the offset
distribution for short GRBs with optical and/or X-ray positions, and
accounting for incompleteness due to bursts with only $\gamma$-ray
positions, is broadly consistent with the predictions of NS-NS merger
models \citep{bsp99,fwh99,bpb+06}.  In particular, we concluded that
$\gtrsim 25\%$ of all short bursts have $\delta R\lesssim 10$ kpc,
while $\gtrsim 5\%$ have offsets of $\gtrsim 20$ kpc.  We repeat the
analysis here using the offsets inferred in \S\ref{sec:kick} for both
the faint galaxies at small angular separations and the brighter
galaxies at larger separations.  As shown in Figure~\ref{fig:offset4},
the model predictions have a median of about 6 kpc, compared to about
4 kpc for the observed sample.  On the other hand, the models predict
$10-20\%$ of offsets to be $\gtrsim 30$ kpc, in good agreement with
the observed distribution in both the $\sim 15$ kpc and $\sim 40$ kpc
scenarios.  We note that the overall smaller offsets measured from the
data may be due to projections effects.  Indeed, the mean correction
factor of $\pi/2$ nicely reconciles the theoretical and observed
distributions.

in the previous section we noted a bimodality in the physical and
host-normalized offsets for {\it Sample 1} and {\it Sample 2}
(Figures~\ref{fig:offset2} and \ref{fig:offset3}).  In the framework
of NS-NS binary kicks this bimodality may indicate that the binaries
generally remain bound to their host galaxies, thereby spending most
of their time at the maximal distance defined by $d_{\rm max}=2GM_{\rm
host}/v_{\rm kick}^2$ (i.e., with their kinetic energy stored as
potential energy; \citealt{bpc+07}).  This would require typical kick
velocities of less than a few hundred km s$^{-1}$.

We further compare the observed offset distribution to predictions for
dynamically-formed NS-NS binaries in globular clusters, with a range
of host galaxy virial masses of $5\times 10^{10}-10^{12}$ M$_\odot$
\citep{sdc+10}.  These models predict a range of only $\approx 5-40\%$
of all NS-NS mergers to occur within 10 kpc of the host center, in
contrast to the observed distribution with about $70\%$ with $\delta
R\lesssim 10$ kpc.  We stress that this result is independent of what
offsets we assign to the bursts in {\it Sample 2} since they account
for only 1/4 of the bursts with optical afterglows.  On the other
hand, the globular cluster origin may account for the bimodality in
the physical and host-normalized offsets (Figures~\ref{fig:offset2}
and \ref{fig:offset3}), with the objects in {\it Sample 2} arising in
globular clusters and the objects with coincident hosts arising from
primordial NS-NS binaries.  This possibility also agrees with the
predicted fraction of dynamically-formed NS-NS binaries of $\sim
10-30\%$ \citep{gpm06}.  The cumulative offset distributions for {\it
Sample 2} alone (assuming the hosts are the galaxies with the lowest
probability of chance coincidence) is well-matched by the range of
predictions for dynamically-formed NS-NS binaries in globular clusters
(Figure~\ref{fig:offset4}).  In this scenario, however, the
implication is that short GRBs outside of globular clusters do not
experience kicks as expected for NS-NS binaries since the largest
measured offset is only 15 kpc.

Unless the populations of short GRBs with only X-ray or $\gamma$-ray
positions have fundamentally different offset distributions, we
conclude that the measured offsets of short GRBs and the predicted
offsets for NS-NS kicks are in good agreement, {\it if we treat all
short GRBs with optical afterglows as a single population}.
Alternatively, it is possible that the bimodal distributions of
physical and host-normalized offsets point to a progenitor bimodality,
with the bursts in {\it Sample 2} originating in globular clusters.

\subsection{Circumburst Densities}
\label{sec:n}

Large offsets are expected to result in low circumburst densities, and
we now investigate whether the observed optical magnitudes for {\it
Sample 2} can be used to place meaningful constraints on the offsets.
The optical afterglow magnitudes do not provide a direct measure of
the circumburst density since they depend on a complex combination of
density, energy, and fractions of the burst energy imparted to the
radiating electrons ($\epsilon_e$) and magnetic fields ($\epsilon_B$).
From our comparison of the X-ray and optical afterglows
(\S\ref{sec:ag}) we were unable to clearly locate the cooling
frequency, which in principle can provide additional constraints on
the density.  However, we can still gain some insight into the
required circumburst densities using a few basic assumptions.  We
assume that the $\gamma$-ray energy ($E_{\rm\gamma ,iso}$) is a
reasonable proxy for the total energy, that $\epsilon_e,
\epsilon_B<1/3$, and that $p=2.2$.  We further use a fiducial redshift
of $z=0.5$ (Figure~\ref{fig:mags}) and a fiducial observation time of
8 hr (\S\ref{sec:ag}).

To determine $E_{\rm\gamma ,iso}$ we use the $15-150$ keV fluences
measured by {\it Swift} (Table~\ref{tab:shbs}), and determine a
correction factor to account for the incomplete energy coverage using
bursts that have also been detected by satellites with a broader
energy range (Figure~\ref{fig:t90_fg}).  We find a typical correction
factor of $\approx 5$ for the $\approx 10-10^3$ keV range.  For the
fiducial redshift of $z=0.5$, the mean energy for short GRBs with
optical afterglows is $E_{\rm\gamma ,iso}\approx 2\times 10^{51}$ erg,
when we include the correction factor (Figure~\ref{fig:z_egiso}).

With these assumptions we find the following relation between the
optical brightness and the circumburst density \citep{gs02}:
\begin{equation}
F_{\rm\nu ,opt}<1\,\,{\rm mJy}\,\times n^{1/2}.
\end{equation}
Given the typical observed fluxes of about 2 $\mu$Jy for {\it Sample
1}, and about 0.6 $\mu$Jy for {\it Sample 2}, we infer typical minimum
densities of $\gtrsim 4\times 10^{-6}$ and $\gtrsim 4\times 10^{-7}$
cm$^{-3}$, respectively.  Thus, the observed optical brightnesses can
be produced even at very low densities that are typical of the IGM.
This indicates that we cannot rule out the large offset scenario based
on density arguments.  Indeed, even if we allow for more typical
values of $\epsilon_e\approx \epsilon_B\approx 0.1$, the resulting
densities are $\approx 4\times 10^{-4}$ and $4\times 10^{-5}$
cm$^{-3}$, respectively.  These results compare favorably with
predictions for NS-NS kicks, which suggest that most mergers will
occur at densities of $\gtrsim 10^{-6}$ cm$^{-3}$ \citep{pb02,bpb+06}.
The distribution of densities in globular clusters is not well known,
preventing a meaningful comparison to our inferred minimum densities
\citep{sdc+10}.

\subsection{Energetics}
\label{sec:e}

\begin{figure}
\epsscale{1}
\plotone{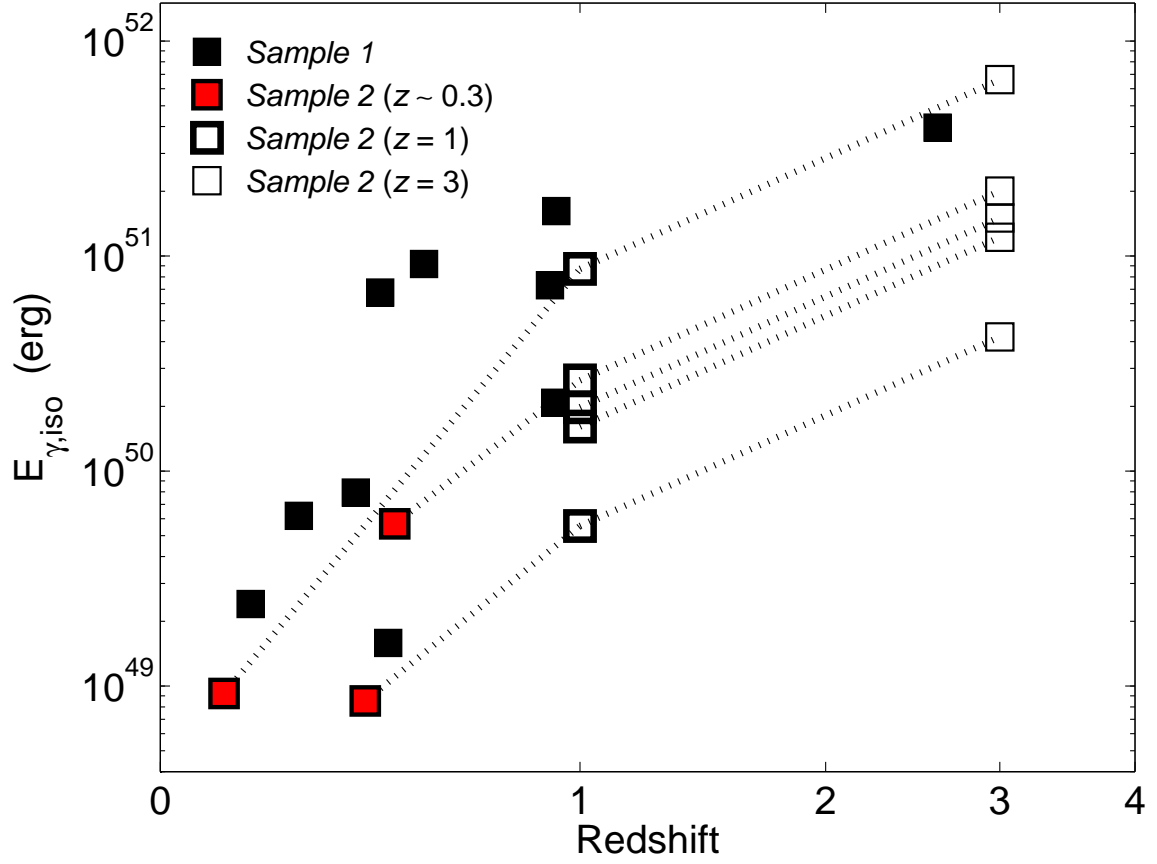}
\caption{Isotropic-equivalent $\gamma$-ray energy as a function of
redshift for the bursts with detected optical afterglows.  We plot the
inferred energies for {\it Sample 2} at the redshifts corresponding to
the galaxies with the lowest probability of chance coincidence (red
squares); at $z\sim 1$ corresponding to the faint galaxies with the
smallest angular separations (thick open squares); and at $z\sim 3$
corresponding to the case of undetected underlying hosts with a
luminosity of about L$^*$ (thin open squares).  The bursts in {\it
Sample 2} match the distribution for the {\it Sample 1} bursts at any
of these redshift intervals.
\label{fig:z_egiso}} 
\end{figure}

Finally, we investigate the energetics of the bursts in {\it Sample 2}
in the context of a high-redshift origin ($z\gtrsim 3$), an origin in
the faint galaxies at $\approx 2\arcsec$ separations ($z\sim 1$), and
an origin in the galaxies with the lowest probabilities of chance
coincidence ($z\approx 0.1-0.5$).  The resulting values of
$E_{\rm\gamma ,iso}$ are shown in Figure~\ref{fig:z_egiso}.  For the
lowest redshift origin, the inferred values are $\approx (1-5)\times
10^{49}$ erg, for a $z\sim 1$ origin they are $\approx 5\times
10^{49}-10^{51}$ erg, and for a $z\gtrsim 3$ origin they are $\approx
5\times 10^{50}-5\times 10^{51}$ erg.

For comparison, the mean value for the bursts in {\it Sample 1} is
$\langle E_{\rm\gamma ,iso}\rangle\approx 8\times 10^{50}$ erg, with a
range of $\approx 10^{49}-5\times 10^{51}$ erg.  However, there is a
clear redshift dependence for the measured $E_{\rm\gamma ,iso}$ values
(Figure~\ref{fig:z_egiso}), with $\approx 5\times 10^{49}$ erg at
$z\lesssim 0.5$ and $\approx 5\times 10^{50}$ erg at $z\gtrsim 0.5$.
Thus, the bursts in {\it Sample 2} generally fit within the known
range of isotropic energies regardless of their actual redshift.
Indeed, at $z\sim 3$ they exhibit similar values of $E_{\rm\gamma
,iso}$ to that of GRB\,090426.  As a result, we cannot use the
inferred energy release as a clear discriminant of the redshift range
for {\it Sample 2}.

If the bursts in {\it Sample 2} indeed originated at high redshifts,
the resulting isotropic-equivalent energies suggest that there is
either a large spread in the energy release of short GRBs (at least 2
orders of magnitude) or a large variations in the ejecta geometry.  If
the typical energy for short GRBs is about $5\times 10^{49}$ erg, as
indicated by the nearest events\footnotemark\footnotetext{We ignore
the factor of $\sim 5$ correction to the energy, due to incomplete
energy coverage (\S\ref{sec:n}), since this would roughly apply to all
bursts in roughly the same way.  The higher redshift events may
require somewhat smaller corrections due to redshifting of the
spectral peak into the softer energy band covered by {\it Swift}.},
then the inferred values of $E_{\rm\gamma ,iso}\approx 2\times
10^{51}$ at $z\sim 3$ indicate opening angles of $\approx
10-15^\circ$.  This is similar to the opening angle of about $7^\circ$
that was inferred for GRB\,051221 \citep{bgc+06,sbk+06}.

\section{Conclusions}
\label{sec:conc}

We undertook the first systematic study of short GRBs with detected
optical afterglows (and hence sub-arcsecond position) but no
coincident host galaxies to limits of $r_{\rm AB}\gtrsim 26$ mag.  We
find that the optical afterglows of these bursts are fainter by about
1.4 mag compared to the optical afterglows of short GRBs with
coincident hosts.  They are similarly fainter in X-rays, and have
somewhat lower $\gamma$-ray fluences and slightly shorter durations.
Both samples have similar ratios of X-ray to optical flux, which are
moreover similar to the ratios measured for long GRBs.  The fainter
afterglows of the bursts lacking coincident hosts may be due to lower
densities, lower energies, or higher redshifts.  However, we note that
only the scenarios with lower energies or higher redshifts naturally
explain the faintness of the prompt emission {\it in the context of
the standard GRB model}.  This is because in the context of internal
processes (shocks or magnetic dissipation) the external density should
not play a role.

We further use deep optical imaging to determine the probability of
chance coincidence for galaxies in the field around each burst, and to
place redshift limits for underlying hosts under the assumption that
they are drawn from the same distribution of the detected short GRB
hosts ($L\approx 0.1-1$ L$^*$).  This analysis leads to the following
possible scenarios: (i) the underlying hosts are fainter than $\sim
26$ mag, indicative of redshifts of $z\gtrsim 1.5$ (if $L\sim 0.1$
L$^*$) or $z\gtrsim 3$ (if $L\sim$\,L$^*$); or (ii) the hosts are
galaxies with substantial offsets --- either faint galaxies at
separations of $\approx 2\arcsec$ ($\approx 15$ kpc for $z\gtrsim 1$)
or brighter galaxies, which we find to be located at $z\sim 0.1-0.5$,
with offsets of $\approx 5-15\arcsec$ ($\approx 30-75$ kpc).  In the
former scenario, unless the galaxies have $L\sim 0.1$ L$^*$, the
resulting redshift distribution is bimodal with peaks at $z\sim 0.5$
and $\sim 3$.  Such a scenario most likely requires a bimodal age
distribution in the context of NS-NS mergers, or two distinct
progenitor systems dominating at low and high redshifts.  While this
cannot be ruled out by present data, the lack of overlap at $z\sim
1-2$ is difficult to explain.  We also conclude that it is unlikely
that any faint underlying hosts are low-luminosity galaxies
($L\lesssim 0.01$ L$^*$) at similar redshifts to the detected hosts,
since this does not naturally explain the difference in afterglow and
prompt emission properties.

In the context of the large offsets scenario, the probability of
chance coincidence (both individually and for the sample) is lower for
the brighter galaxies at offsets of $\approx 5-15\arcsec$ than for the
faint galaxies at offsets of $\approx 2\arcsec$.  However, it is not
unlikely that the true associations are a mix of both populations,
since in each case there is a non-negligible probability that $1-2$ of
the associations are spurious.  We note that optical positions are
crucial for a uniform comparison of the chance coincidence
probabilities for faint galaxies with small offsets and bright
galaxies with larger offsets; in the case of only X-ray positions
there is an inherent bias (in the sense of {\it a posteriori}) against
associations with faint galaxies inside the error circle, despite the
fact that they are consistent with no offset.

From spectroscopic observations for 3 of the 5 bursts we find that the
galaxies with the lowest probability of chance coincidence are a star
forming galaxy at $z=0.111$ (GRB\,061201), an early-type galaxy at
$z=0.473$ (GRB\,070809), and an early-type cluster member galaxy at
$z=0.403$ (GRB\,090515).  If these associations are indeed correct,
they only slightly alter the host demographics, which are dominated by
star forming galaxies \citep{ber09}.  However, this does suggest that
our present understanding of the relative ratios of star forming and
elliptical hosts may be incomplete.

The resulting distributions of angular, physical, and host-normalized
offsets for the bursts with and without coincident hosts appear to be
distinct, rather than continuous.  However, the joint distribution of
projected physical offsets is in good agreement with theoretical
predictions for NS-NS binary mergers.  On the other hand, the
predicted distribution for dynamically-formed NS-NS binaries in
globular clusters provides a much poorer fit to the entire data set,
unless they account for only the bursts with large offsets ({\it
Sample 2}).  In the case of a hybrid population of primordial and
dynamically-formed binaries, with the latter accounting for only the
large offsets, the fact that all the remaining offsets ({\it Sample
1}) are $\lesssim 10$ kpc, is indicative of no significant kicks.  The
large physical offsets also naturally explain the fainter afterglow
emission as a result of lower circumburst densities.  The resulting
isotropic $\gamma$-ray energies match the observed distribution for
short GRBs with coincident hosts, either at $z\sim 0.3$ or at
$z\gtrsim 1$.

Our conclusion that large offsets of $\sim 15-70$ kpc (corresponding
to $\approx 4-10$ galactic effective radii) are a likely explanation
for the bursts with optical afterglows and no coincident hosts is of
fundamental importance.  The only progenitor model that naturally
explains this result in the merger of NS-NS/NS-BH binaries, most
likely due to kicks, or possibly with a minor contribution from a
globular cluster population (accounting specifically for the events in
{\it Sample 2}).  While a conclusive demonstration of a large offset
requires an absorption redshift measurement that matches an offset
galaxy redshift, the distribution of optical afterglow magnitudes
indicates that this will be difficult to achieve.  Indeed, an
absorption redshift is available for only one likely short GRB
\citep{lbb+10}.

\begin{figure}
\epsscale{1}
\plotone{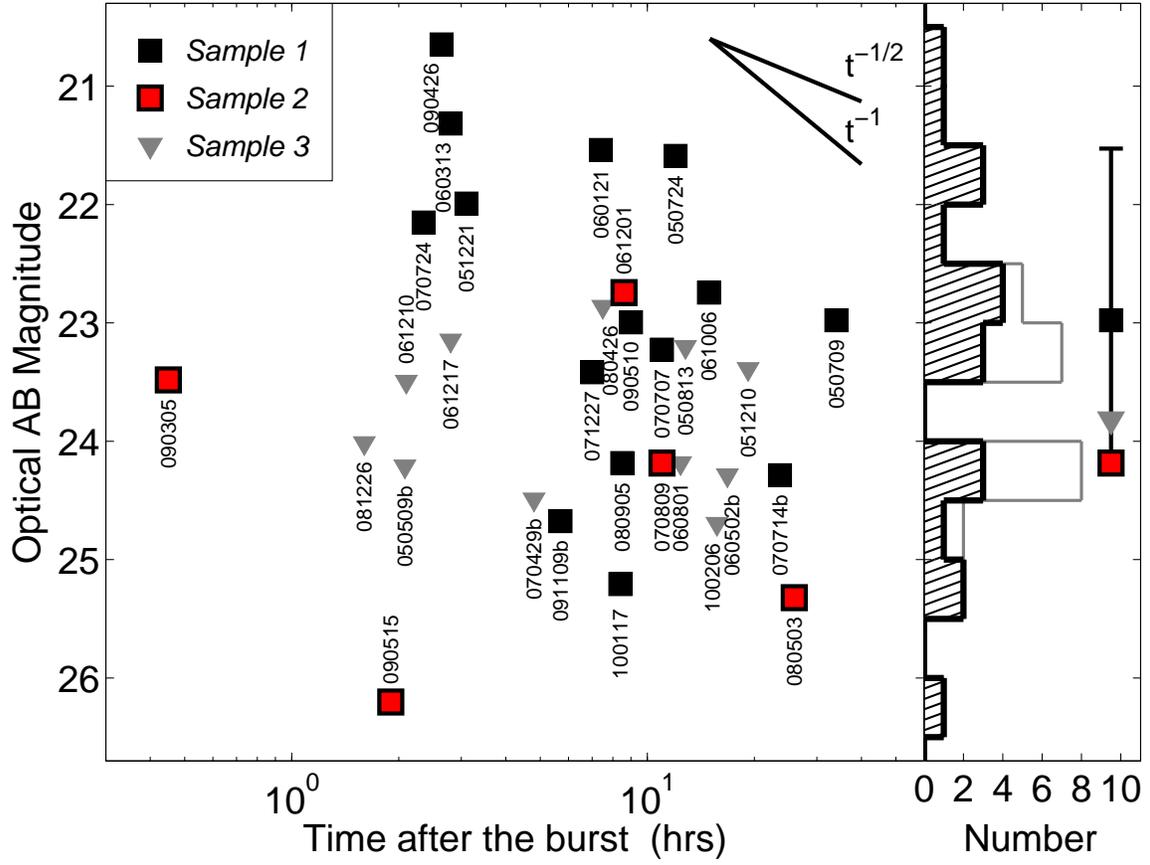}
\caption{Optical afterglow brightness on timescales of a few hours
after the burst for short GRBs with detected afterglows ({\it Sample
1}: black squares; {\it Sample 2}: red squares) or upper limits (gray
triangles).  The lines at the top right indicate the fading tracks for
afterglow decay rates of $\alpha=-0.5$ and $-1$.  The right panel
shows the projected histogram for the bursts with detected afterglows
(hatched) and upper limits (open).  The symbols mark the mean for each
sample, and the vertical bar marks the standard deviation for {\it
Sample 1}.
\label{fig:optag1}} 
\end{figure}

\begin{figure}
\epsscale{1}
\plotone{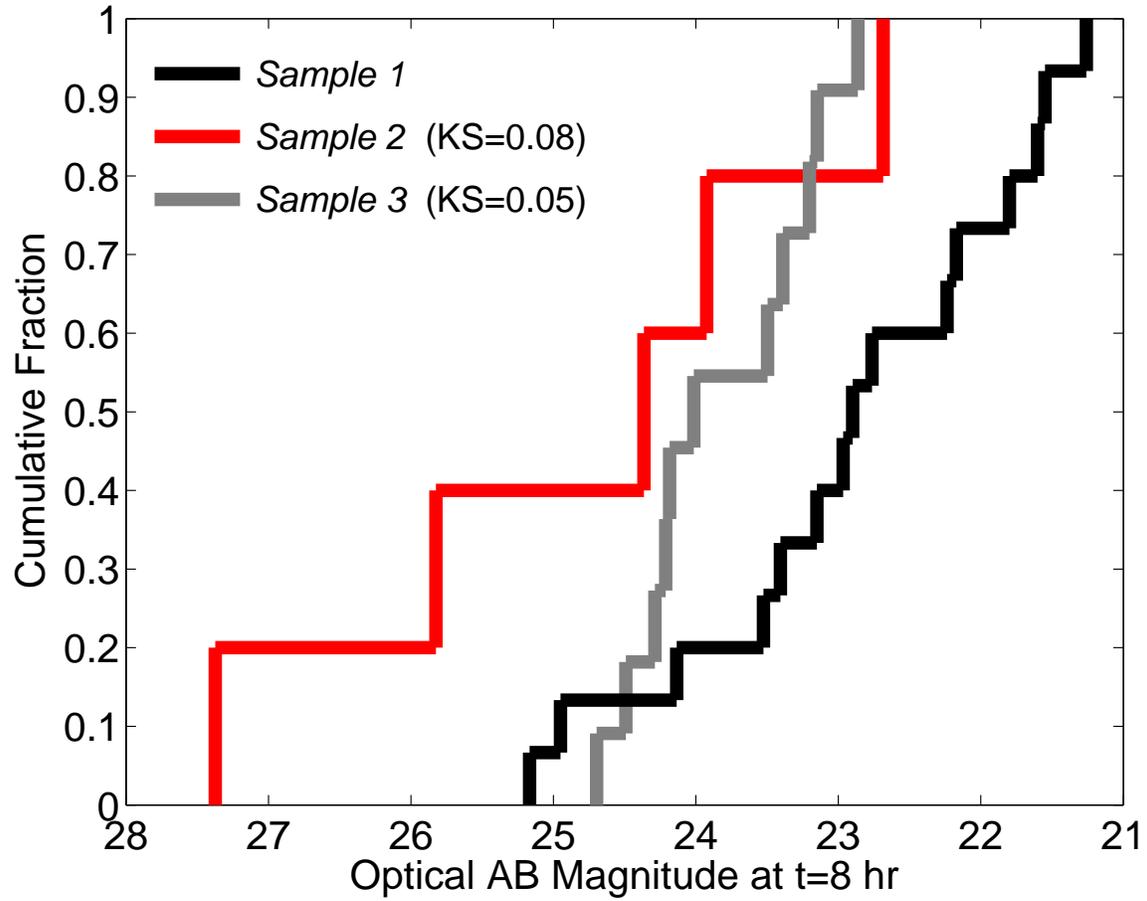}
\caption{Cumulative afterglow brightness distributions for the bursts
in Figure~\ref{fig:optag1}, extrapolated to a common fiducial time of
8 hr after the burst with a fading rate of $\alpha=-0.75$.  The K-S
probabilities relative to the sample with detected afterglows and
coincident hosts are noted in the figure.  It appears unlikely that
the bursts with no coincident hosts, and those with deep upper limits,
are drawn from the same distribution as the bursts with detected
hosts.
\label{fig:optag2}} 
\end{figure}

\begin{figure}
\epsscale{1}
\plotone{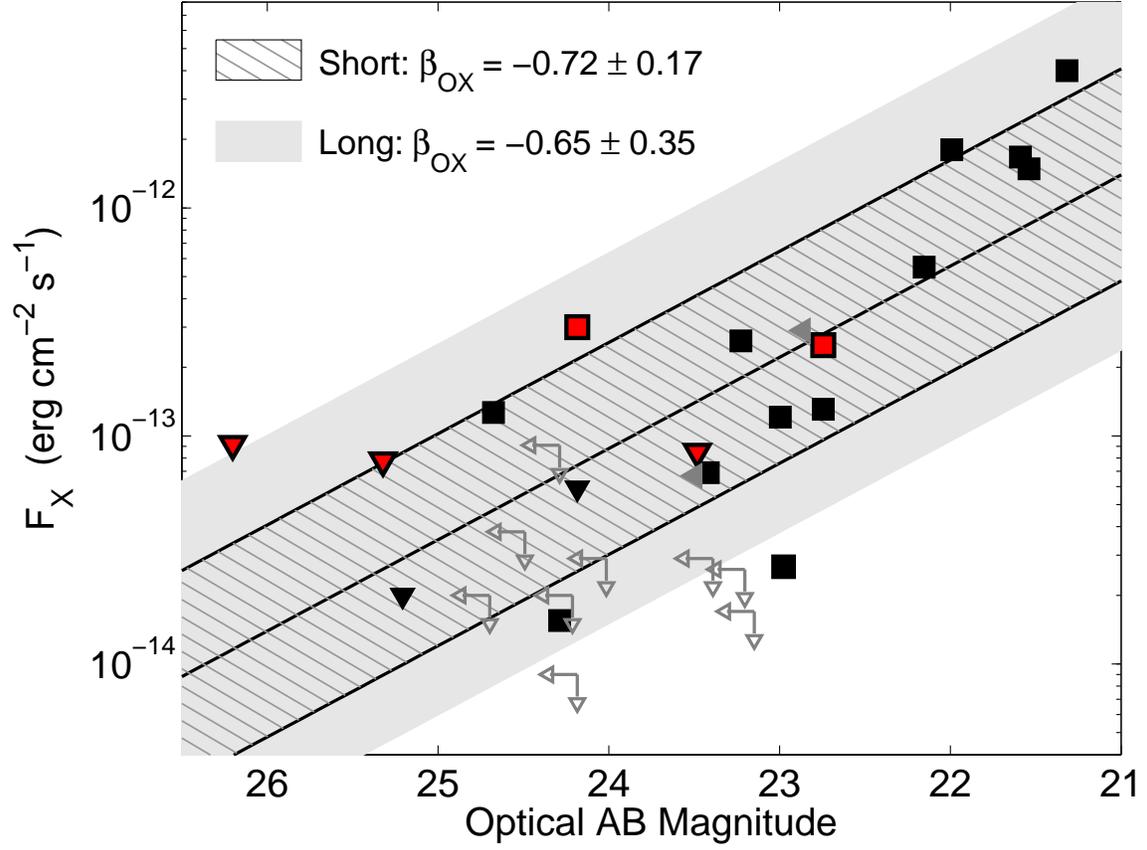}
\caption{X-ray versus optical flux for the bursts from
Figure~\ref{fig:optag1}.  The cross-hatched region marks the median
and standard deviation assuming the expected power-law correlation
with an index $\beta_{\rm OX}$.  The light shaded region marks the
region occupied by long GRBs \citep{jhf+04}.  The distributions for
long and short GRBs are largely indistinguishable, as are the
distributions for short GRBs with an without coincident hosts.  We
note that a large fraction of the bursts with optical upper limits
also have undetected X-ray afterglows on timescales of a few hours
after the burst.  The overall similarity between the ratio of optical
to X-ray flux for long and short GRBs does not allow us to clearly
locate the synchrotron cooling frequency ($\nu_c$) in relation to the
X-ray band.  If $\nu_c>nu_X$ for short GRBs, the faintness of the
optical afterglows for bursts with no coincident hosts cannot be used
to distinguish density and redshift effects.
\label{fig:opt_xray}} 
\end{figure}

We end with several important observations.  First, rapid optical
follow-up of short GRBs with 8-meter class telescopes is essential
since observations to a depth of about 25 mag within a few hours after
the burst may recover nearly all optical afterglows, regardless of
circumburst density (Figure~\ref{fig:optag1}).  Second, short GRBs
with only X-ray positions are unlikely to provide strong support for
either negligible or large offsets due to the appreciable size of the
error circles and the fact that the sample with optical afterglows
exhibits wide host galaxy diversity, i.e., events with coincident
bright hosts at $z\sim 0.2-1$, events with coincident faint hosts
likely at $z\gtrsim 1$, and events with no coincident hosts likely due
to offsets.  Third, a meaningful study of short GRB offsets requires a
statistical approach to mitigate the shortcomings of {\it a
posteriori} chance coincidence probabilities, as well as to
incorporate the relevant information from afterglow and prompt
emission observations.

Finally, we note that from a wide range of observations of both the
afterglows and host galaxies it appears that the case for NS-NS
mergers as the progenitors of short GRBs is gaining observational
support.  Our main result here is that short GRBs with optical
afterglows and no detected host galaxies are somewhat less likely to
be explained with high redshifts or with dwarf galaxy hosts at low
redshift.  Instead, these bursts likely exploded $\sim 15$ kpc from
galaxies at $z\sim 1$ or tens of kpc from galaxies at $z\sim 0.3$.
With larger samples, these possibilities will allow us to improve our
understanding of the short GRB redshift distribution, the host galaxy
demographics, and predictions for gravitational wave detections.  If
high redshifts indeed turn out to be prevalent, this will have a
significant effect on the possibility of multiple progenitor
populations.

\acknowledgements We acknowledge helpful discussions with Alicia
Soderberg, Ryan Chornock, and Josh Grindlay.  This work was partially
supported by Swift AO5 grant number 5080010, and made use of data
supplied by the UK Swift Science Data Centre at the University of
Leicester.

\clearpage
\begin{deluxetable}{lcccccccl}
\tabletypesize{\scriptsize}
\tablecolumns{9}
\tabcolsep0.08in\footnotesize
\tablewidth{0pc}
\tablecaption{Properties of Short GRBs with Optical Afterglows or
Limits
\label{tab:shbs}}
\tablehead {
\colhead {GRB}                  &
\colhead {$T_{90}$}             &
\colhead {$z\,^a$}              &
\colhead {$F_\gamma\,^b$}       &
\colhead {$t_X$}                &
\colhead {$F_X\,^c$}            &
\colhead {$t_{\rm opt}$}        &
\colhead {$F_{\rm\nu, opt}$}    &
\colhead {Refs.}    \\
\colhead {}                     &
\colhead {(s)}                  &
\colhead {}                     &
\colhead {(erg cm$^{-2}$)}      &
\colhead {(hr)}                 &
\colhead {(erg cm$^{-2}$ s$^{-1}$)} &
\colhead {(hr)}                 &
\colhead {($\mu$Jy)}            &
\colhead {}                     
}
\startdata
\multicolumn{8}{c}{\normalsize Short GRBs with Optical Detections ({\it Samples 1} \& {\it 2})} \\\hline
050709  & 0.07 & 0.161                  & $2.9\times 10^{-7}$ & 60.5 & $3.5\times 10^{-15}$  & 34   & 2.3   & 1--3 \\ 
050724  & 3.0  & 0.257                  & $3.9\times 10^{-7}$ & 11.8 & $1.7\times 10^{-12}$  & 12   & 8.4   & 4--6 \\ 
051221A & 1.4  & 0.546                  & $1.2\times 10^{-6}$ & 3.1  & $1.8\times 10^{-12}$  & 3.1  & 5.8   & 7--8 \\ 
060121  & 2.0  & $<4.3$                 & $4.7\times 10^{-6}$ & 6.1  & $1.8\times 10^{-12}$  & 7.4  & 8.8   & 9--11 \\ 
060313  & 0.70 & $<3$                   & $1.1\times 10^{-6}$ & 2.8  & $4.0\times 10^{-12}$  & 2.8  & 10.8  & 11--12 \\ 
061006  & 0.42 & 0.438                  & $1.4\times 10^{-6}$ & 19.5 & $1.0\times 10^{-13}$  & 14.9 & 2.9   & 11,13 \\ 
061201$\,^d$ & 0.80 & $<1.7$,\,\,0.111? & $3.3\times 10^{-7}$ & 8.6  & $2.5\times 10^{-13}$  & 8.6  & 2.9   & 14--15 \\ 
070707  & 1.1  & $<4.3$                 & $1.4\times 10^{-6}$ & 12.5 & $2.3\times 10^{-13}$  & 11   & 1.9   & 16 \\ 
070714B & 3.0  & 0.923                  & $7.2\times 10^{-7}$ & 24.4 & $1.5\times 10^{-14}$  & 23.6 & 0.7   & 17 \\ 
070724  & 0.40 & 0.457                  & $3.0\times 10^{-8}$ & 2.35 & $5.5\times 10^{-13}$  & 2.3  & 5.0   & 18--19 \\ 
070809$\,^d$ & 1.3  & $<3$,\,\,0.473?   & $1.0\times 10^{-7}$ & 11.0 & $3.0\times 10^{-13}$  & 11   & 0.8   & 20--21 \\ 
071227  & 1.8  & 0.381                  & $2.2\times 10^{-7}$ & 7.0  & $6.9\times 10^{-14}$  & 7.0  & 1.6   & 13 \\ 
080503$\,^d$ & 0.32 & $<3$,\,\nod       & $6.1\times 10^{-8}$ & 66.0 & $<7.8\times 10^{-14}$ & 25.9 & 0.3   & 22 \\ 
080905  & 1.0  & 0.122                  & $1.4\times 10^{-7}$ & 18.4 & $<5.9\times 10^{-14}$ & 8.5  & 0.8   & 23 \\ 
090305$\,^d$ & 0.40 & $<4.3$,\,\nod     & $7.5\times 10^{-8}$ & 0.45 & $<8.5\times 10^{-14}$ & 0.45 & 1.5   & 24--25 \\ 
090426  & 1.28 & 2.609                  & $2.5\times 10^{-7}$ & 2.0  & $1.1\times 10^{-12}$  & 2.6  & 20.0  & 26--27 \\ 
090510  & 0.30 & 0.903                  & $3.4\times 10^{-7}$ & 6.4  & $1.7\times 10^{-13}$  & 9.0  & 2.3   & 28 \\ 
090515$\,^d$ & 0.04 & $<4.3$,\,\,0.403? & $2.1\times 10^{-8}$ & 4.9  & $<9.2\times 10^{-14}$ & 1.9  & 0.1   & 29 \\ 
091109B & 0.30 & \nod                   & $1.9\times 10^{-7}$ & 3.8  & $1.9\times 10^{-13}$  & 5.7  & 0.5   & 30 \\ 
100117  & 0.30 & 0.92                   & $9.3\times 10^{-8}$ & 8.4  & $<2.0\times 10^{-14}$ & 8.4  & 0.3   & 31 \\\hline\\
\multicolumn{8}{c}{\normalsize Short GRBs with Optical Limits ({\it Sample 3})} \\\hline
050509B & 0.04 & 0.225         & $9.5\times 10^{-9}$ & 9.1  & $<2.0\times 10^{-14}$ & 2.1  & 0.7 & 32--33 \\ 
050813  & 0.60 & \nod          & $1.2\times 10^{-7}$ & 94.9 & $<2.6\times 10^{-14}$ & 12.8 & 1.9 & 34--36 \\ 
051210  & 1.27 & \nod          & $8.1\times 10^{-8}$ & 6.9  & $<2.9\times 10^{-14}$ & 19.2 & 1.6 & 11,37  \\ 
060502B & 0.09 & \nod          & $4.0\times 10^{-8}$ & 18.6 & $<9.1\times 10^{-14}$ & 16.8 & 0.7 & 11,38  \\ 
060801  & 0.50 & 1.130         & $8.1\times 10^{-8}$ & 10.9 & $<9.0\times 10^{-15}$ & 12.4 & 0.8 & 11   \\ 
061210  & 0.19 & 0.409         & $1.1\times 10^{-6}$ & 72.4 & $<6.7\times 10^{-14}$ & 2.1  & 1.4 & 11   \\ 
061217  & 0.21 & 0.827         & $4.6\times 10^{-8}$ & 38.6 & $<1.7\times 10^{-14}$ & 2.8  & 2.0 & 11   \\ 
070429B & 0.50 & 0.902         & $6.3\times 10^{-8}$ & 11.8 & $<3.8\times 10^{-14}$ & 4.8  & 0.6 & 39   \\ 
080426  & 1.30 & \nod          & $3.7\times 10^{-7}$ & 6.1  & $2.9\times 10^{-13}$  & 7.5  & 2.6 & 40   \\ 
080702  & 0.50 & \nod          & $3.6\times 10^{-8}$ & 2.9  & $<4.0\times 10^{-14}$ & 12.1 & 12. & 41   \\ 
081226  & 0.40 & \nod          & $9.9\times 10^{-8}$ & 3.2  & $<2.9\times 10^{-14}$ & 1.6  & 0.9 & 42   \\ 
100206  & 0.12 & \nod          & $1.4\times 10^{-7}$ & 6.9  & $<2.0\times 10^{-14}$ & 15.7 & 0.5 & 43               
\enddata 
\tablecomments{Prompt emission and afterglow data for short GRBs with
detected optical afterglows (top section) and deep optical afterglow
limits (bottom section).\\
$^a$ Redshifts include spectroscopic measurements, limits from
afterglow detections in the UV/optical, and for the bursts in {\it
Sample 2}, redshifts for galaxies with the lowest probability of
chance coincidence (marked by ?).\\
$^b$ The fluences are in the observed $15-150$ keV band, with the
exception of GRB\,050709 ($2-400$ keV) and GRB\,060121 ($2-400$
keV).\\
$^c$  All XRT data are from \citet{ebp+07} and \citet{ebp+09}.\\
$^d$ Short GRBs in {\it Sample 2}.\\
References: [1] \citet{vlr+05}; [2] \citet{ffp+05}; [3]
\citet{hwf+05}; [4] \citet{bcb+05}; [5] \citet{bpc+05}; [6]
\citet{gbp+06}; [7] \citet{bgc+06}; [8] \citet{sbk+06}; [9]
\citet{ucg+06}; [10] \citet{ltf+06}; [11] \citet{bfp+07}; [12]
\citet{rvp+06}; [13] \citet{dmc+09}; [14] \citet{sdp+07}; [15]
\citet{fbf10}; [16] \citet{pdc+08}; [17] \citet{gfl+09}; [18]
\citet{bcf+09}; [19] \citet{ktr+10}; [20] \citet{gcn6739}; [21]
\citet{gcn7889}; [22] \citet{pmg+09}; [23] \citet{rwl+10}; [24]
\citet{gcn8933}; [25] \citet{gcn8934}; [26] \citet{adp+09}; [27]
\citet{lbb+10}; [28] \citet{mkr+10}; [29] \citet{row+10}; [30]
\citet{mul+09}; [31] Fong et al.~in prep.; [32] \citet{gso+05}; [33]
\citet{bpp+06}; [34] \citet{fsk+07}; [35] \citet{ber06}; [36]
\citet{pbc+06}; [37] \citet{lmf+06}; [38] \citet{bpc+07}; [39]
\citet{cbn+08}; [40] \citet{gcn7644}; [41] \citet{gcn7977}; [42]
\citet{gcn8736}; [43] \citet{gcn10395}
}
\end{deluxetable}

\clearpage
\begin{deluxetable}{lcccc}
\tabletypesize{\scriptsize}
\tablecolumns{5}
\tabcolsep0.1in\footnotesize
\tablewidth{0pc}
\tablecaption{Observations of Short GRBs with Optical Afterglows and
no Coincident Host Galaxies ({\it Sample 2})
\label{tab:nohost}}
\tablehead {
\colhead {GRB}                  &
\colhead {Instrument}           &
\colhead {Filter}               &
\colhead {$t_{\rm exp}$}        &
\colhead {$m_{\rm lim}\,^a$}    \\
\colhead {}                     &
\colhead {}                     &
\colhead {}                     &
\colhead {(s)}                  &
\colhead {(AB mag)}             
}
\startdata
061201 & HST/ACS        & F814W & 2224 & 26.0 \\
070809 & Magellan/LDSS3 & $r$   & 1500 & 25.4 \\
080503 & HST/WFPC2      & F606W & 4000 & 25.7 \\
090305 & Magellan/LDSS3 & $r$   & 2400 & 25.6 \\
090515 & Gemini-N/GMOS  & $r$   & 1800 & 26.5
\enddata
\tablecomments{$^a$ Limits are $3\sigma$.}
\end{deluxetable}

\clearpage

\clearpage

\clearpage

\clearpage

\clearpage

\clearpage

\clearpage

\clearpage

\clearpage

\clearpage

\clearpage

\clearpage

\clearpage

\clearpage

\clearpage

\clearpage

\clearpage

\clearpage

\clearpage

\clearpage

\clearpage

\end{document}